\documentclass[pre,twocolumn,notitlepage,amsmath,amssymb]{revtex4-1}

\usepackage{graphicx}
\usepackage{float}
\usepackage{placeins}
\usepackage{color, colortbl}
\usepackage{multirow}
\usepackage{array}
\usepackage[overload]{empheq}

\newcolumntype{M}[1]{>{\centering\arraybackslash}m{#1}}

{

\begin{document}

\title{Negative Feedback, Linearity and Parameter Invariance in Linear Electronics}

\author{Luciano da F. Costa$^1$}
\email{ldfcosta@gmail.com}
\author{Filipi N. Silva$^1$}
\author{Cesar H. Comin$^1$}

\affiliation{$^1$S\~ao Carlos Institute of Physics, University of S\~ao Paulo, PO Box 369, 13560-970, S\~ao Carlos, SP, Brazil\\}


\begin{abstract}
Negative feedback is a powerful approach capable of improving several aspects of a system. In linear electronics, it has been critical for allowing invariance to device properties. Negative feedback is also known to enhance linearity in amplification, which is one of the most important foundations of linear electronics. At the same time, thousands of transistors types have been made available, suggesting that these devices, in addition to their known variability of parameters, have distinguishing properties. The current work reports a systematic approach to quantifying the potential of negative feedback, with respect to bipolar transistors, as a means to providing device invariance and linearity. Several methods, including concepts from multivariate statistics and complex systems, are applied at the theoretical as well as experimental levels, and a number of interesting results are obtained and discussed. For instance, it has been verified that the transistors types indeed have well-defined characteristics which clearly segregate them into groups. The addition of feedback at moderate and intense levels promoted uniformization of the properties of these transistors when used in a class A common emitter configuration. However, such effect occurred with different efficiency regarding the considered device features, and even intense feedback was unable to completely eliminate device dependence. This indicates that it would be interesting to consider the device properties in linear design even when negative feedback is applied. We also verified that the linearization induced in the considered experiments is relatively modest, with effects that depend on type of transfer function of the original devices.
\end{abstract}

\maketitle

\section{Introduction}

The beginning of electronics can be traced back to the use of Galen crystals as detectors in the first AM radios (later, these rudimentary diodes would be found to be semiconductors).  Such devices were soon replaced by vacuum tubes, which dominated the evolution of electronics from the 10's to the 60's, when electronics switched back to semiconductors, a situation that extends to the present day.  As such, semiconductors were (and are) crucial for the full development of electronics and the implied revolutions in telecommunications, informatics, and networking that reshaped the modern world. 

The embracing of semiconductors in the 60's was justified, in contrast to vacuum tubes, by their smaller sizes, power dissipation and excellent potential for mass production at low prices~\cite{braun1982revolution}.  Yet, semiconductors had two major shortcomings: influence of temperature and wide variation of constants (or parameters) among devices of the same type.  Thus, the adoption of semiconductors for linear electronics depended critically on means for controlling these two unwanted effects.  This was ultimately accomplished by using circuit configurations capable of stabilizing the influence of temperature and of greatly reducing the effect of  device parameters on the circuit performance.  This was achieved, in great part, thanks to ubiquitous application of negative feedback, an approach pioneered by Harold Black~\cite{black1934stabilized}.  More specifically, in this approach, gain is traded off for device independence and linearity, often accompanied by bandwith extension and noise rejection (e.g.~\cite{chen2016active}).  The viable amount of negative feedback is therefore limited by the available gain, and often more than one amplifying stages are required to compensate for the given up gain. 
Yet, the impressive success of linear electronics is the living proof of the efficiency
of feedback in implementing device parameter independency.  The efficiency of feedback for this finality, and also for improving linearity, is widely acknowledged in the technical and scientific literature (e.g.~\cite{self2013audio,sedra1998microelectronic}).  Yet, few related quantitative, systematic, investigations of interrelationships between feedback, linearity and device invariance are available.  

Strikingly, about 80000 different models of transistors have been developed~\footnote{According to http://alltransistors.com/}.   Even allowing for  different applications (e.g. high frequency, high voltage, etc.),
there seems to be too many device choices given the uniformizing effects of negative feedback.  As a matter of fact, several transistors can be used for the same finality (e.g. audio) which, again, suggests variability of performance.  The relative lack of more systematic approaches to quantifying the effects of negative feedback on such a varied universe of semiconductor devices, especially transistors, motivated the current approach.

While a completely comprehensive investigation is beyond the scope of
this work, we approached the above problem by considering devices sampled from several
popular small signal bipolar junction transistor (BJT) types, mainly for audio applications. Theoretical and experimentally derived data regarding their individual characteristics (including indicators of linearity) when used in class A amplifiers were then obtained and analysed in the absence and presence of negative feedback.  Then, by using multivariate statistical~\cite{joliffe1992principal,johnson2002applied} and pattern recognition~\cite{bishop2006pattern,witten2005data} concepts and methods, we were able to derive several conclusions driving interesting implications and possibilities for new approaches regarding the effects of negative feedback for circuit design.   We observe that the potential of such results is particularly strong because linear amplification constitutes the core of linear electronics~\cite{sedra1998microelectronic}, with a wide range of applications in  instrumentation, telecommunications, audio/hifi, biomedicine, microelectronics, and control.  Moreover, the obtained results are directly extensible to other areas, such as neuroscience, biophysics, ecology, climate, psychology, mechanics, chemistry, complex systems and networks, among many other possibilities.

The list of results and findings reported in the current work is ample, being presented and respectively discussed in the main body of the article.  A summary of the main presented results includes the identification of consistent and differentiating electronics properties for most types of considered small signal BJTs (considering a representative set of measurements and parameters characterizing resistances, gains, and linearity of the devices and circuits), the finding that two linear combinations of such variables are capable of expressing the greatest part of the variation of the properties of the considered devices, the confirmation of the effect of moderate and intense negative feedback in reducing the dependence on device parameters (leading to overlapping groups of devices), the identification of relatively moderate effect of negative feedback on the circuit linearity, the demonstration of critical variation of negative feedback on linearization depending on the type of non-linearity (with many interesting peculiarities regarding the attenuation of induced harmonics in terms of the feedback gain), the identification that the efficiency of feedback in reducing device dependence varies with the different types of circuit properties (with voltage gain being the most resilient), the consequent identification of the fact that even intense negative feedback is not completely capable of erasing the memory of the circuit with respect to original device properties, and a quantitative study of how sensitive circuit parameters (with emphasis on voltage gain) can be with respect to device properties and values of the resistors in the considered circuit.   Such results bear several implications for circuit design, and are immediately extensible to other types of devices (e.g. FETs, MOS, vacuum tubes), circuit configurations, as well as other areas.

The present text is organized as follows.   We start by presenting the several adopted main concepts and methods (which include multivariate statistics and pattern recognition approaches), and proceeds to presenting, in respective order, the theoretical and experimental investigations performed, obtained results, and discussions.  A list of possible further works is presented in the concluding sections.

\section{Basic concepts}
\label{s:basicConcepts}

In this section we present the main adopted concepts, including analogue aspects of transistors, the transfer function of a device, the class A amplifier configuration, principles of negative feedback, total harmonic distortion, multivariate statistical and pattern recognition concepts and methods for data analysis.

\subsection{Negative feedback}
\label{s:negFeedback}

\begin{figure}[]
  \begin{center}
  \includegraphics[width=\linewidth]{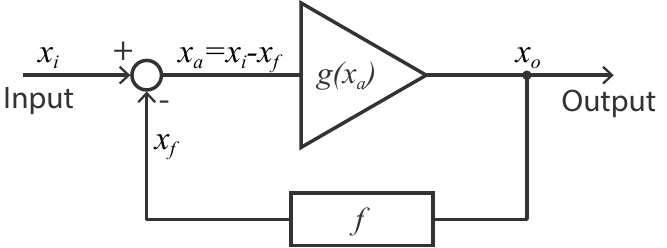} \\
  \caption{Illustration of negative feedback. An input signal is amplified according to a transfer function $g(x_a)$, and a fraction $f$ of the result is subtracted from the input.}
  ~\label{f:feedbackBlocks}
  \end{center}
\end{figure}

The typical operation of feedback in an amplifier is shown in Figure~\ref{f:feedbackBlocks}. A feedback signal $x_f$ 
is subtracted from the input signal $x_i$. The result (or `error'), $x_a$, is amplified according to a transfer function $g(x_a)$, 
defining the output signal $x_o=g(x_a)$. This output signal is then used as feedback $x_f=fx_o$ for the input signal~\cite{chen2016active}. 
Therefore, it follows that $x_o=g(xi-fx_o)$. The total amplification of the circuit with feedback can be written as

\begin{align}
A^f & = \frac{x_o}{x_i} \nonumber \\
    & = \frac{g(x_a)}{x_a+fg(x_a)}.\label{eq:ampGeneralGeneralFeedback}
\end{align}
For a linear transfer function, $g(x_a)=Ax_a$ and the amplification is given by

\begin{equation}
A^f = \frac{A}{1+Af}.\label{eq:ampGeneralFeedback}
\end{equation}

\begin{figure}[]
  \begin{center}
  \includegraphics[width=0.8\linewidth]{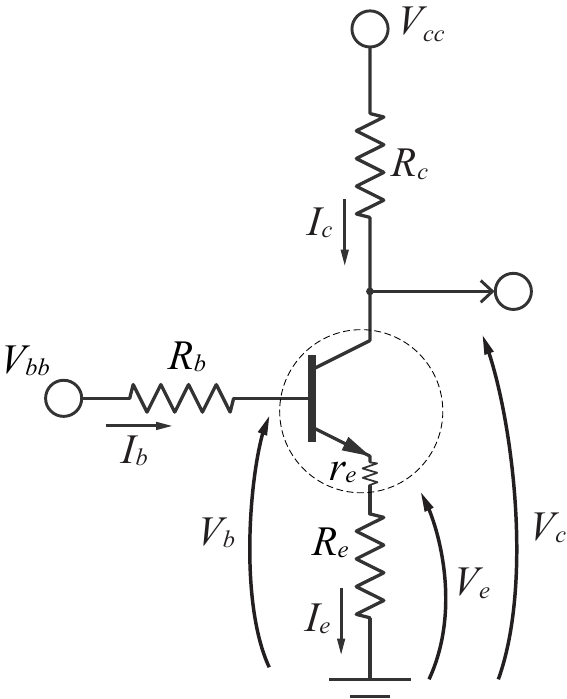} \\
  \caption{Class A common emitter amplifier circuit used in this work. }
  ~\label{f:amplifierWithFeedback}
  \end{center}
\end{figure}

In Figure~\ref{f:amplifierWithFeedback} we show the common emitter amplifier configuration used in the current work.  The total amplification of this circuit is defined as

\begin{equation}
A_c^f=\frac{V_c}{V_{bb}}.\label{eq:AcfDefMainText}
\end{equation}
In Appendix A we show how to write $A_c^f$ as a function of the circuit resistances and transistor properties, which results in 

\begin{equation}
A_c^f \approx\frac{R_c\beta r_o}{R_b(r_o - R_c)+\beta r_o(R_e+r_e)}. \label{eq:AcfApproxMainText}
\end{equation}
Therefore, the amplification of the circuit with feedback is influenced by $\beta$ and $r_o$. In the absence of feedback, $R_e=0$, and the amplification is

\begin{equation}
A_c \approx\frac{R_c\beta r_o}{R_b(r_o - R_c)+\beta r_o r_e}. \label{eq:AcApproxMainText}
\end{equation}

We can associate Equation~\ref{eq:AcfApproxMainText} with the general expression of a feedback circuit, provided by Equation~\ref{eq:ampGeneralFeedback}. This is done by replacing $A^f$ and $A$ in Equation~\ref{eq:ampGeneralFeedback} by their respective values, $A_c^f$ and $A_c$, for the considered circuit. In appendix A we show that this results in

\begin{equation}
f \approx \frac{R_e}{R_c}.
\end{equation}
Value $f$ (also known as feedback gain) represents the fraction of output returned as input to the transistor. The actual amount of feedback used in the circuit is given by $fA_c$, that is,

\begin{equation}
fA_c \approx \frac{R_e\beta r_o}{R_b(r_o - R_c)+\beta r_o r_e} \label{eq:feedbackApproxMainText}
\end{equation}
Thus, increasing $R_e$ or $R_c$, or decreasing $R_b$, lead to higher values of feedback. 

Other properties of the circuit, such as the output and input resistances, can also be analytically obtained. In addition, we can also calculate the sensitivity of the voltage gain with respect to different properties of the transistor and the circuit. The sensitivity of a variable $y$ relative to the variation of another variable $x$ is defined as~\cite{chen2016active}

\begin{equation}
S_y(x) = \frac{x}{y}\frac{dy}{dx}.
\end{equation}
In Appendix A we derive equations for the sensitivity of the gain $A_c^f$ with respect to $\beta$, $r_o$ and $A_c$. In Table~\ref{t:teoEquations} we present a summary of all obtained equations relating device and circuit parameters.  

\begin{table*}[]
\centering
\begin{tabular}{ll}
\toprule
Voltage gain & $\displaystyle A_c^f \approx\frac{R_c\beta r_o}{\mathcal{\tilde{D}}}$ \\
Open loop gain & $\displaystyle A_c \approx\frac{R_c\beta r_o}{R_b \left(r_o-R_c\right)+\beta r_o r_e}$ \\
Feedback return & $\displaystyle f \approx \frac{R_e}{R_c}$ \\
Output resistance & $\displaystyle R_{o} \approx r_o\frac{R_b + \beta R_e + \beta r_e}{R_b+R_e+r_e}$ \\
Input resistance & $\displaystyle R_{i} \approx r_o \frac{R_b + \beta R_e+\beta r_e}{R_e+r_e+r_o}$ \\
Gain sensitivity with $\beta$ & $\displaystyle S_{A_c^f}(\beta)\approx \frac{\left(R_b+R_e+r_e\right) \left(R_e-R_c+r_o+r_e\right)}{\mathcal{\tilde{D}}}$ \\
Gain sensitivity with $r_o$ & $\displaystyle S_{A_c^f}(r_o) \approx \frac{\left(R_b+R_e+r_e\right) \left(R_e-R_c+r_e\right)}{\mathcal{\tilde{D}}}$ \\
Gain sensit. with open gain & $\displaystyle S_{A_c^f}(A_c) \approx \frac{r_e \beta r_o+R_b \left(r_o-R_c\right)}{\mathcal{\tilde{D}}}$ \\
\hline
\multicolumn{2}{l}{$\mathcal{\tilde{D}}=R_b \left(r_o-R_c\right)+\beta r_o(R_e + r_e)$} \\
\end{tabular}
\caption{Main equations obtained in the theoretical analysis of negative feedback in class A common emitter amplifiers.}
\label{t:teoEquations}
\end{table*}

\subsection{Transfer functions}

\begin{figure}[]
  \begin{center}
  \includegraphics[width=\linewidth]{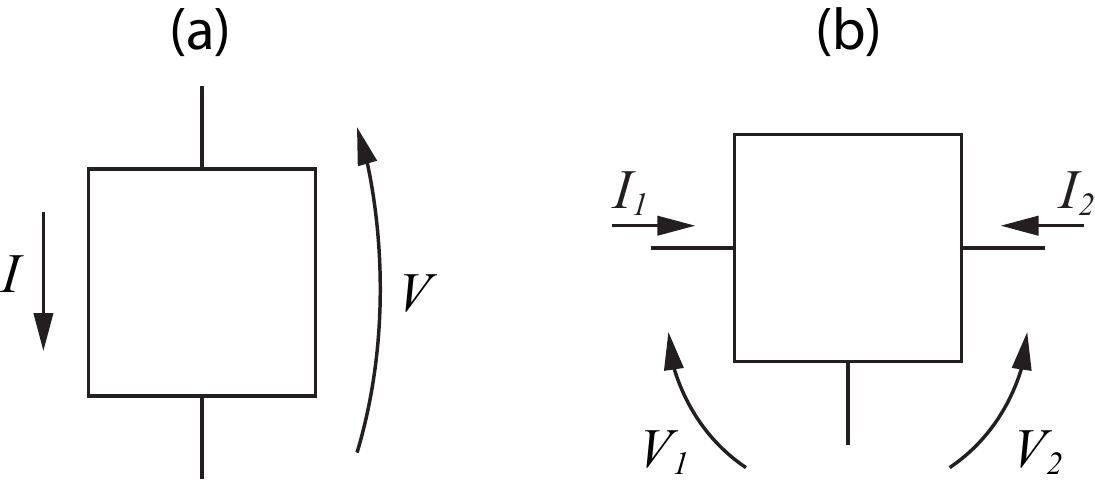} \\
  \caption{Two- (a) and three-terminal (b) configurations typically used to model
     relationships between properties in electrical and electronic devices.}
  ~\label{f:portModels}
  \end{center}
\end{figure}

Given a (linear or non-linear) device represented as an analog system, 
its operation can be understood in terms of its characteristic curve and surface 
relating input and output. In the current work, we use the 2- and 3-terminal 
representations of devices, as illustrated in Figure~\ref{f:portModels}.  The former can be used to represent the relationship between electrical quantities (e.g. current and voltage) of a 2-terminal device, and the latter is inherently suitable to represent amplifying devices (e.g. transistors, vacuum tubes) and circuits (e.g. the common emitter class A amplifying stage).

\begin{figure}[]
  \begin{center}
  \includegraphics[width=0.6\linewidth]{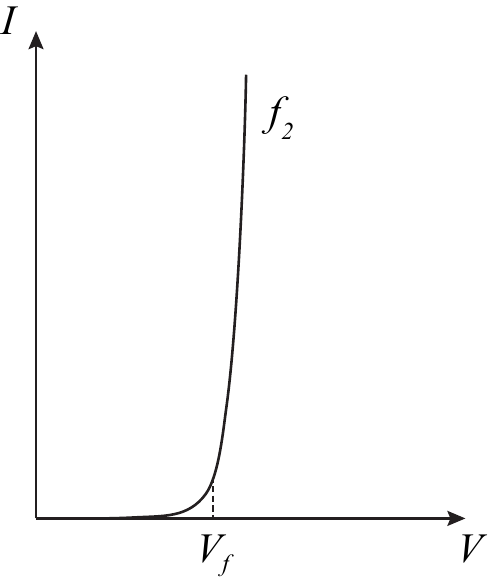} \\
  \caption{Generic transfer function of a Silicon diode, expressing current
    in terms of voltage, i.e. $I=f_2(V)$.}
  ~\label{f:diodeCurve}
  \end{center}
\end{figure}

In the case of the 2-terminal representation (shown in Figure~\ref{f:portModels}(a)),
 we can relate current $I$ and voltage $V$ through the expressions $V = f_1(I)$ or $I=f_2(V)$, where $f_1$ and $f_2$ are denominated \emph{transfer functions}.  Figure~\ref{f:diodeCurve} shows the kind of transfer function $f_2$ typically obtained 
for Silicon diodes.

The output behaviour of 3-terminal devices (illustrated in Figure~\ref{f:portModels}(b)), 
i.e. $V_2$ or $I_2$, can be expressed by using any of the following four alternative relationships~\cite{gray1990analysis, sedra1998microelectronic,chen2016active}:

\begin{align}
V_2 & = f_z(I_1, I_2) \nonumber\\
V_2 & = f_g(V_1, I_2) \nonumber\\  
I_2 & = f_h(I_1, V_2) \nonumber\\
I_2 & = f_y(V_1, V_2) \nonumber
\end{align}

Each of these mappings can be understood as a two-dimensional scalar field, which can be
restricted by imposing some constraint relating the two output variables (e.g. a load line $I_2 = g(V_2)$), giving rise to a respective \emph{transfer function} directly relating the output variable in terms of just one of the input variables (e.g. $V_2 = f_g(V_1,g(V_2))$ in the case of the previous example).  The next section provides examples of such transfer functions.

\subsection{Transistors}

\begin{figure}[]
  \begin{center}
  \includegraphics[width=0.4\linewidth]{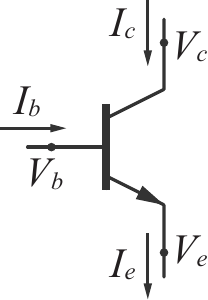} \\
  \caption{The traditionally adopted symbol for an NPN BJT, and the conventions
  for currents and voltages considering the junctions BE and BC to be directly and
  inversely biased, respectively.}
  ~\label{f:transistorBasics}
  \end{center}
\end{figure}

Transistors are 3-terminal (emitter, base, collector) devices aimed at switching 
or linear applications~\cite{sedra1998microelectronic}.  Bipolar junction devices (BJTs) can be of NPN or PNP types, but in the current work we will be limited to the former case.  
Typically, in amplifying configurations the base-emitter junction is forward biased, while the base-collector junction is reversely biased.  Figure~\ref{f:transistorBasics} shows an NPN BJT and the currents and  voltages involved.   The current gain, $\beta$ or $h_{fe}$, of the transistor is  defined as

\begin{equation}
\beta=\frac{I_c}{I_b}\label{eq:beta}.
\end{equation}
Typically, $\beta$ is large (e.g. 300). The output resistance, $r_o$, is given
by

\begin{equation}
r_o=\frac{V_c}{I_c}\label{eq:ro}.
\end{equation}
A low value of this parameter if often desired.

A more complete understanding of the linear operation of an NPN transistor configured as above involves the consideration of transfer functions and hybrid small signal relationships, as briefly reviewed in Appendix A.

\subsection{Early voltage}
\label{s:EarlyExplanation}

\begin{figure}[]
  \begin{center}
  \includegraphics[width=\linewidth]{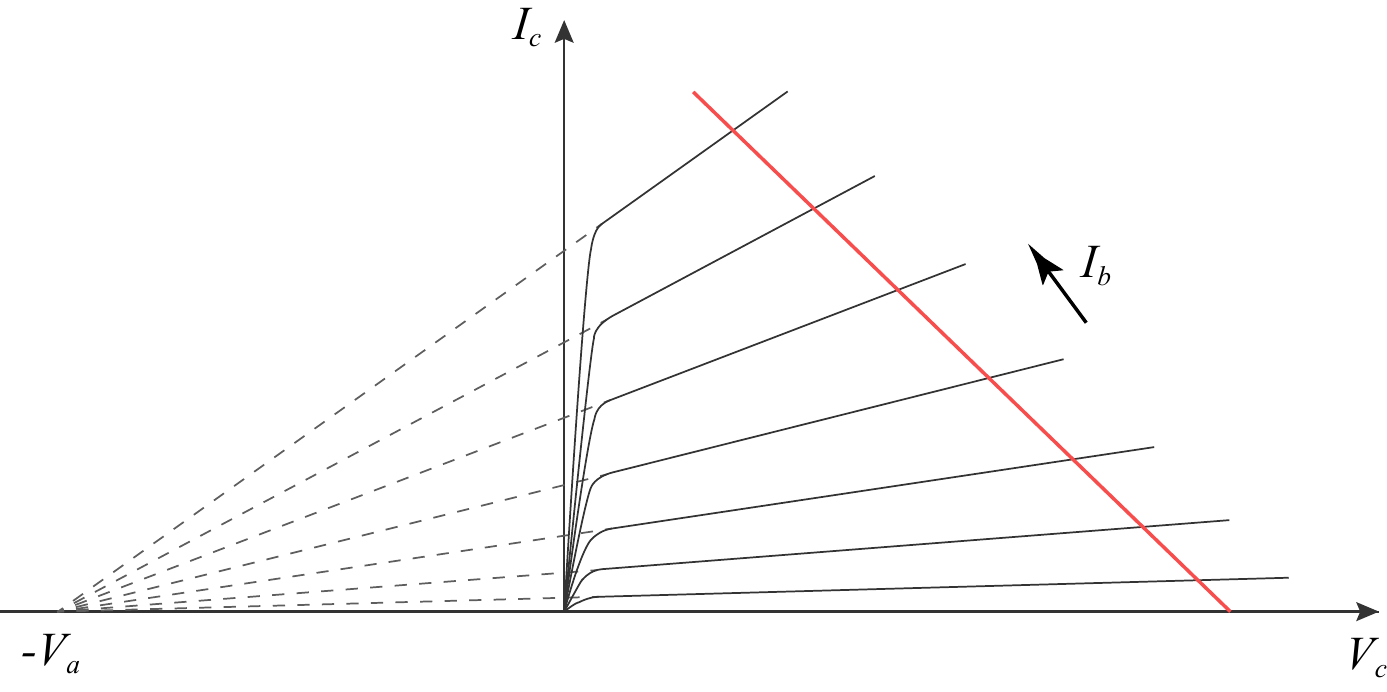} \\
  \caption{Generic representation of the $I_c \times V_c \times I_b$ characteristic 
  curves and their relationship with the Early voltage.  Please refer to the text
  for explanation and discussion.}
  ~\label{f:earlyVoltageIlust}
  \end{center}
\end{figure}

Another interesting parameter of the transistor is its \emph{Early voltage}~\cite{jaeger1997microelectronic}, henceforth represented as $V_a$.  As illustrated in Figure~\ref{f:earlyVoltageIlust}, this voltage corresponds to the intersection of the more linear portions of the characteristic curves of the device for various $I_b$ parameters.  This figure also includes a generic load line L, shown in red, which constrains the operation of a transistor (e.g. in the common emitter configuration without feedback, such a line completely defines the current and voltage relationship at the output).  It is interesting to observe that the smaller the absolute value of $V_a$, the larger will be the variation of $I_b$ along different parts of the load line.  More specifically, in the lower portion of this line, $I_b$ tends to vary slower than at the upper parts.  This clearly implies in a non-linear relationship between input ($I_b$) and output ($I_c$ or $V_c$).  Therefore, the Early voltage is closely related to the non-linearity of each given device, constituting an important parameter to be taken into account.  Appendix B presents a numeric procedure for estimating $V_a$.

\subsection{Class A common emitter amplifiers}

\begin{figure}[]
  \begin{center}
  \includegraphics[width=0.8\linewidth]{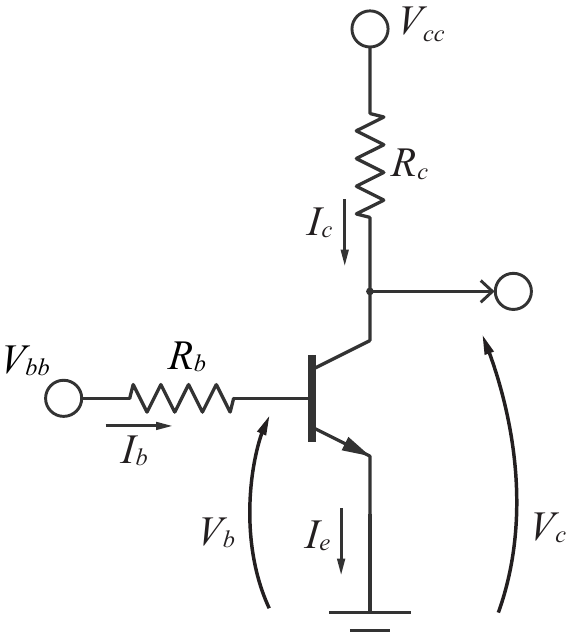} \\
  \caption{Example of class A common emitter amplifier without feedback.}
  ~\label{f:amplifierWithoutFeedback}
  \end{center}
\end{figure}

Amplifiers can be classified in several categories depending on their type
of operation.  In the configuration known as \emph{class A}, the input signal
is treated as a whole (i.e. both positive and negative parts are kept together) 
by the amplifier, which requires considerable power dissipation even in the absence of signal (null input).  Class A amplifiers are compatible with a minimalist approach, which often tends to promote linearity, e.g. by avoiding the crossover potentially implied by class AB designs. Figure~\ref{f:amplifierWithoutFeedback} shows a single stage BJT class A amplifier in common emitter configuration without feedback.

\begin{figure}[]
  \begin{center}
  \includegraphics[width=\linewidth]{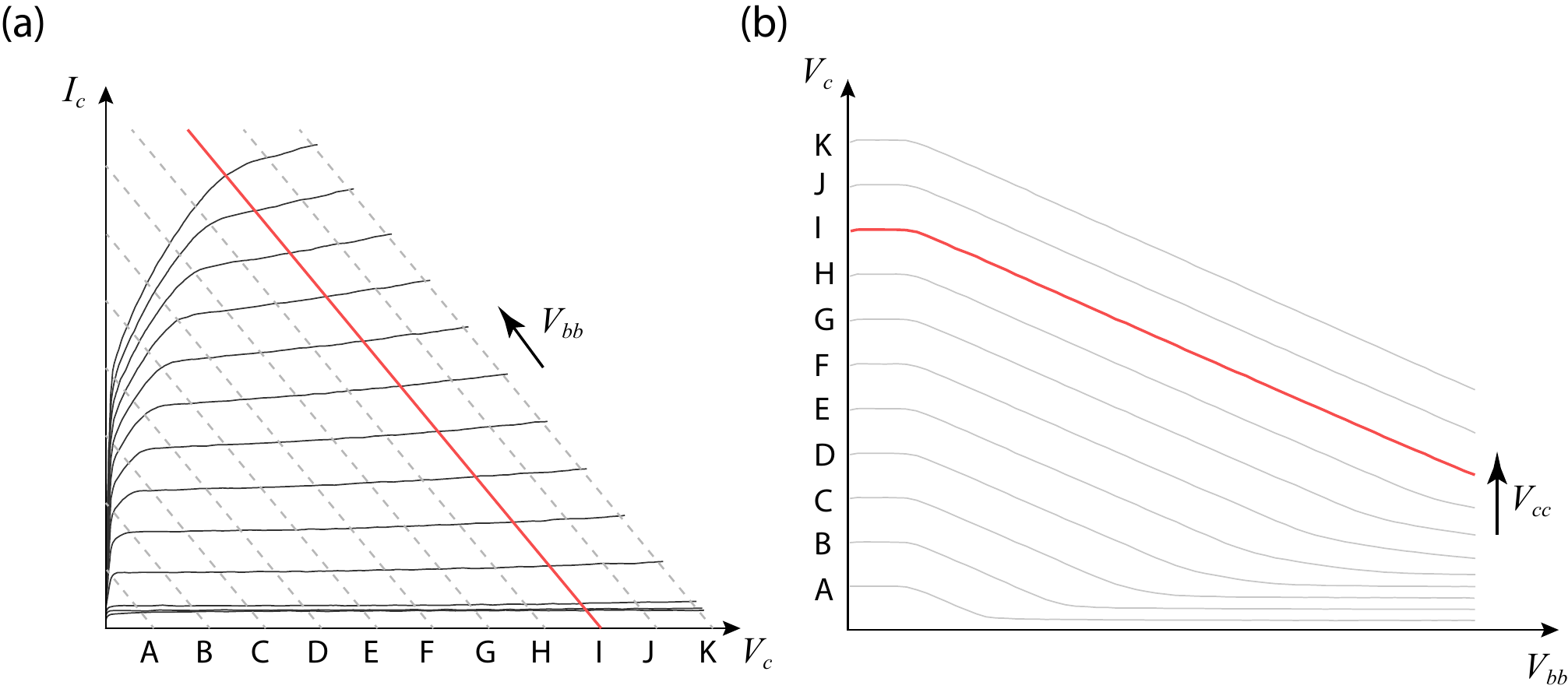} \\
  \caption{ (a) Typical characteristic curves (isolines) for a transistor, associating output current $I_c$ and voltage $V_c$ for given values of input voltage $V_{bb}$. Examples of load lines are shown in dashed and red lines labeled from A to K. (b) Transfer curves associated to the load lines shown in (a). }
  ~\label{f:isoTransf}
  \end{center}
\end{figure}

The operation of the amplifier can be described by three main variables, $I_c$, $V_c$ and $V_{bb}$, where $V_{bb}$ and $V_c$ are, respectively, the input and output voltages, and $I_c$ is the collector current. The relationship between these variables defines a surface, $S(I_c, V_c, V_{bb})$, in the $I_c \times V_c \times V_{bb}$ space. Such a surface is usually represented by isolines, or characteristic curves, of specific values of $V_{bb}$ in a two-dimensional space defined by variables $I_c$ and $V_c$, as shown in Figure~\ref{f:isoTransf}(a). The resistance $R_c$ and the main power supply $V_{cc}$ of the amplifier define a load line for the operation of the circuit. In Figure~\ref{f:isoTransf}(a) examples of load lines are represented as gray dashed lines. The relationship between the input and output voltages for a given load line is called the \emph{transfer function} of that load line. Examples of transfer functions for distinct load lines are illustrated in Figure~\ref{f:isoTransf}(b). The red line indicates the transfer function associated with the load line indicated in Figure~\ref{f:isoTransf}(a).

\subsection{Total harmonic distortion -- THD}

A traditional way to study the linearity of an amplifier is by estimating its total 
harmonic distortion (THD)~\cite{cordell2011designing}. For a given frequency $f$, this 
measurement can be obtained by using a pure sinusoidal function with frequency $f$ as 
input, identifying new harmonic components in the output (a perfectly linear amplifier 
would produce no such components), and taking the ratio between the magnitudes of these 
spurious harmonics ($V_{2f}$, $V_{3f}$, etc) and of the fundamental ($V_f$). More 
formally, the THD can be calculated as:

\begin{equation}
THD(f) = {\sqrt{V_{2f}^2 + V_{3f}^2 + V_{4f}^2 + \cdots} \over V_f}\label{eq:THDDef}
\end{equation}

Because the load is purely resistive, the same THD will be attained irrespectively of 
the input frequency $f$. Therefore, we considered a sinusoidal function with $f=1kHz$.

\subsection{Principal component analysis and scatter distance}

\begin{figure}[]
  \begin{center}
  \includegraphics[width=0.8\linewidth]{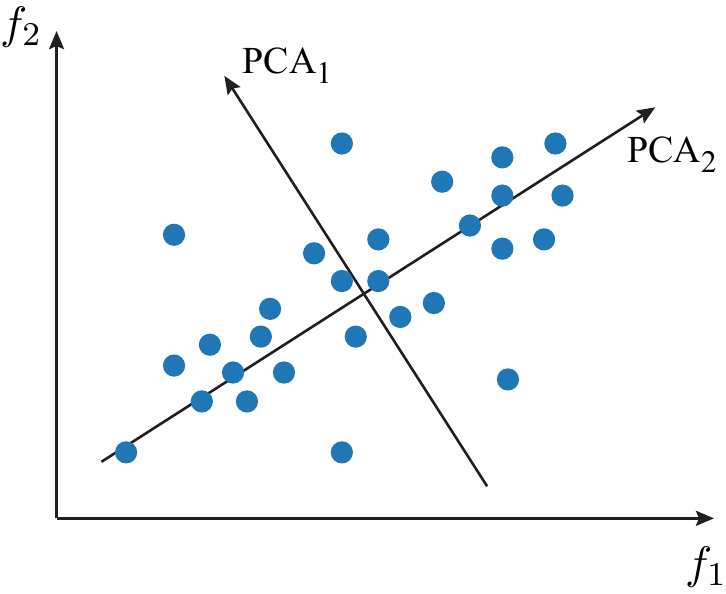} \\
  \caption{Illustration of PCA application to a 2-dimensional dataset.}
  ~\label{f:PCAIllustration}
  \end{center}
\end{figure}

The interpretation and modeling of datasets containing large number of features is
a difficult task~\cite{bishop2006pattern}. A common approach for simplifying the complexity of the data and making it easier to interpret is to project the original $m$-dimensional data (where $m$ is the number of features) into a 2-dimensional space. Many techniques have been
proposed for such a purpose~\cite{fodor2002survey}, including Principal Component Analysis (PCA)~\cite{joliffe1992principal}. PCA 
is widely used for defining a linear projection based on the maximum variance 
of the original data. For instance, suppose that the original data contains 2 features, $f_1$
and $f_2$, as shown in Figure~\ref{f:PCAIllustration}. PCA can then be used for defining new axes $\mathrm{PCA}_1$ and $\mathrm{PCA}_2$, 
where $\mathrm{PCA}_1$, called the first axis, follows the direction of maximum variance of the 
data. If $\mathrm{PCA}_1$ contains enough information about the data, that is, the variance
along the $\mathrm{PCA}_1$ axis is much larger than along $\mathrm{PCA}_2$, the latter axis can be removed and
the data can now be represented by the single variable $\mathrm{PCA}_1$.

More generally, given an $m$-dimensional space, PCA can be used to define $m$ new variables, 
$\mathrm{PCA}_1,\mathrm{PCA}_2,\dots,\mathrm{PCA}_m$, where $\mathrm{PCA}_1$ contains the most variance about the data, 
$\mathrm{PCA}_2$ is the second most informative and so on. For visualization purposes, it is common to 
consider only the $\mathrm{PCA}_1$ and $\mathrm{PCA}_2$ axes in the analysis. The amount of variance retained by
each $\mathrm{PCA}_i$ axis is indicated by the normalized $i$-th eigenvalue of the covariance matrix of the original 
data~\cite{joliffe1992principal}, here represented as $E_i$. This quantity is defined in the range $[0\%,100\%]$, that is, $E_1=100\%$ means that the first PCA axis contains all information about the data, while lower values of $E_1$
indicates a loss of information. Also, each PCA axis can be analyzed in terms of its respective weights. This is because each
element $w_i$ defining the new axis defines how much of the respect $i$-th measurement is contained in that axis. For instance, suppose that a new PCA axis applied to a dataset containing 4 variables has weights $W=(2.3, 0.1, 0.4, 2.1)$. This means that this axis is mainly a linear combination of the first and fourth variables, while the second and third ones are mostly irrelevant for the axis. 

When the dataset contains classes, that is, different categories for the objects, a common criterion for quantifying the separation between the classes is based on the so-called \emph{scatter matrix}~\cite{fukunaga2013introduction,cios1998data}. This matrix contains the distances between pairs of classes normalized by the variance of data inside each class. The trace of this matrix, henceforth referred as \emph{scatter distance}, can then be used for quantifying the overall separation between the classes.

\section{Theoretical analysis}
\label{s:theoreticalAnalysis}

This section addresses a theoretical/symbolic analysis of the linearity and device invariance properties of negative feedback.  

\subsection{Transfer function linearity and THD}

\begin{figure*}[]
  \begin{center}
  \includegraphics[width=\linewidth]{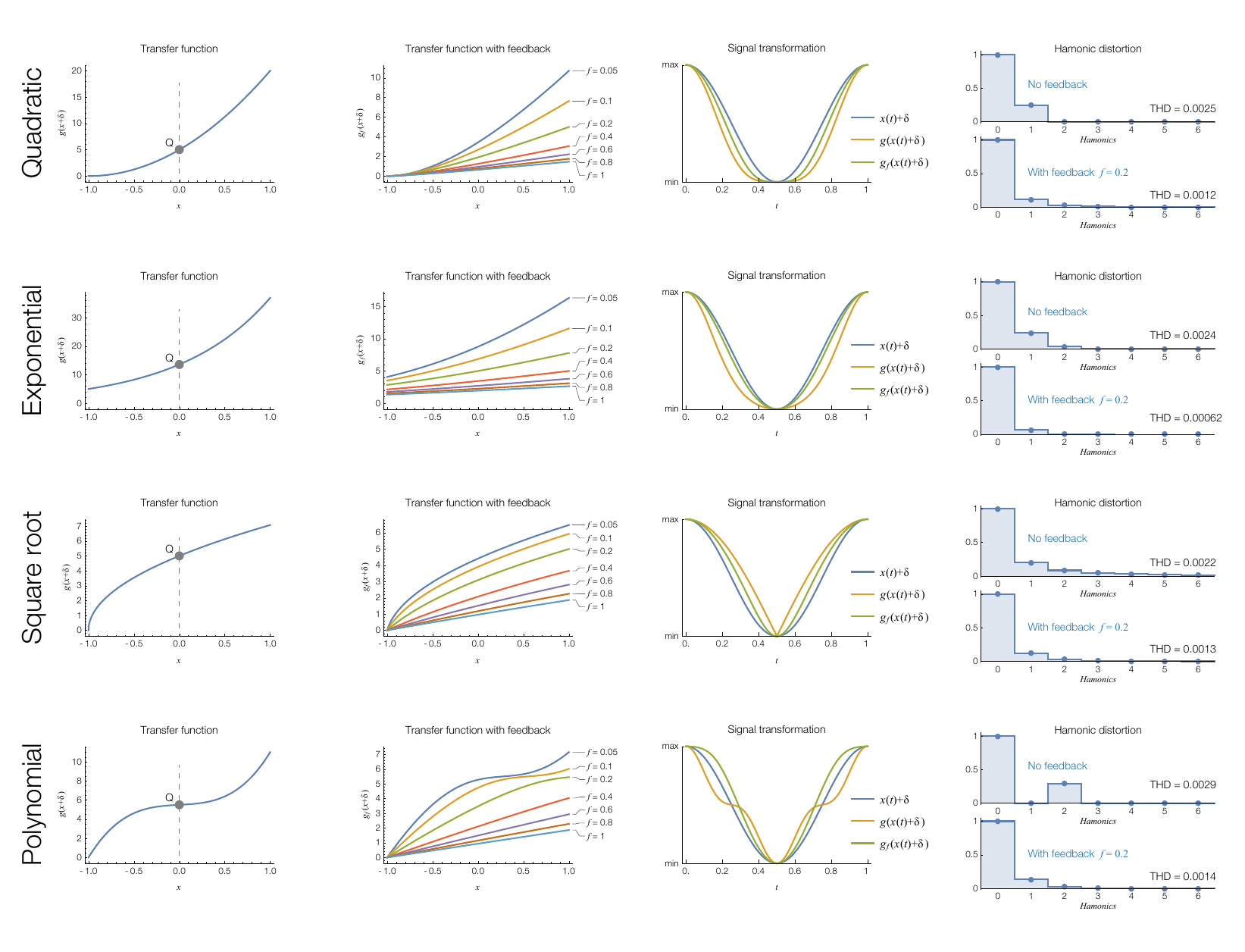} \\
  \caption{The considered non-linear transfer functions (firs column) and the respectively linearized versions obtained by negative feedback (second column), the input and output cosine waves (third column), and the harmonics distribution (fourth column). The operation points used for THD calculation are indicated as grey dots in each plot of the first column. The open loop gain, $A$, used in the analysis is $A=5$.}
  ~\label{f:theoreticalFeedbackAndTHD}
  \end{center}
\end{figure*}

Insights about the effect of feedback on linearization of transfer functions can be gained by performing a preliminary analytical investigation of some representative non-linear cases (given in Table~\ref{t:transferFunctions}), which is done in this section.  We chose the quadratic, exponential, square root and a third-degree polynomial transfer functions in order to provide a representative set of possible non-linearities.  For instance, the quadratic function is believed to provide an approximation of the transfer functions of triodes (vacuum tubes) (e.g.~\cite{zolzer2002dafx}), while the exponential transfer function is a good model of the behaviour of BJTs (e.g.\cite{Williams1991analog}). Symbolic methods were applied to obtain the effective transfer functions in presence of various degrees of negative feedback, quantified in terms of $f$.  We also obtained, analytically, the respective THDs, so as to quantify the achieved degree of linearity.  The results are presented in Figure~\ref{f:theoreticalFeedbackAndTHD}, which shows the several transfer functions with the respectively chosen operation points (first column) and the effective transfer functions obtained through feedback for several values of $f$ (second column).   The third column shows, for each respective case, an input cosine (blue) function and the respective outputs in absence (orange) and presence of negative feedback (green) for $f=0.2$.  The last column presents the relative  magnitude of harmonics impinged on a pure cosine without (upper plot) and with feedback (lower plot) for each considered transfer function, together with the respective THD values.  Observe that the first bar in the frequency graphs corresponds to the fundamental ($100\%$), which is included for comparison purposes.

\begin{table}[]
\centering
\begin{tabular}{ll}
\toprule
Function name	& Equation \\
\hline
Quadratic & $g = Ax^2$ \\
Square root & $g = A\sqrt{x}$ \\
Exponential & $g = Ae^x$ \\
Third-degree polynomial & $g = A[(x-1)^3 + \frac{x}{10} + 1]$ \\
\hline
\end{tabular}
\caption{Non-linear transfer functions considered in the theoretical analysis of THD.}
\label{t:transferFunctions}
\end{table}

The linearization ability of feedback as $f$ increases is clearly visible in the second column of Figure~\ref{f:theoreticalFeedbackAndTHD}.  In all cases, the effective transfer function produced by feedback tends steadily to a straight line as $f$ is increased.  Noticeably, even the changes of concavity in the polynomial transfer function (last row), are reduced by feedback as it acts in order to promote a more linear response, at least as it is visually perceived.

Interesting results were obtained regarding the harmonic alterations induced on a cosine input by each of the considered non-linear transfer functions (please refer to the third and fourth columns in Figure~\ref{f:theoreticalFeedbackAndTHD}).  First, it is interesting to recall that no harmonics would be added by a perfectly linear systems, so that a flat, null, distribution of added harmonic magnitudes would be therefore obtained.  Indeed, the incorporation of harmonics is an intrinsic property of non-linear systems.  As is clear from the fourth column in Figure~\ref{f:theoreticalFeedbackAndTHD}, the distributions of harmonics vary substantially with respect to each type of non-linearity.  As expected~\cite{self2013audio}, the quadratic function introduces a strong first harmonic component, with unnoticeable effects on higher harmonics.  The incorporation of feedback reduces considerably the first harmonic distortion but, at the same time and interestingly, incorporates a second harmonic component to the cosine transformation, which brings up the point that negative feedback is, ultimately, a non-linear operation that promotes linearity!  Actually, the overal reduction of the respective THDs is only about half.  The exponential non-linearity impinges all orders of harmonics to the cosine, and all were attenuated in the situation considered here, with a substantial overall reduction of the THD.  The square root transfer function constitutes another interesting case.  First, among the considered cases, it corresponds to the one that induces more substantial harmonic components.  Second, the distribution of such harmonics is such that the feedback cannot greatly promote linearity, resulting in a THD that is only about half of the original transfer function, without feedback.  The last considered non-linear transfer function corresponds to a third-degree polynomial containing all terms.  It has been motivated by the type of non-linearity found in push-pull amplifiers~\cite{self2013audio}, characterized by a low-gain transition region.  At least for the configuration considered here, a strong second harmonic component is added by this transfer function in the absence of feedback.  This harmonic is greatly attenuated by feedback, but at the expense of having as a remainder a significant first harmonic component.  The difficulties of feedback in linearizing this case are reflected in the relatively small overall THD reduction.

\begin{figure*}[h]
  \begin{center}
  \includegraphics[width=0.7\linewidth]{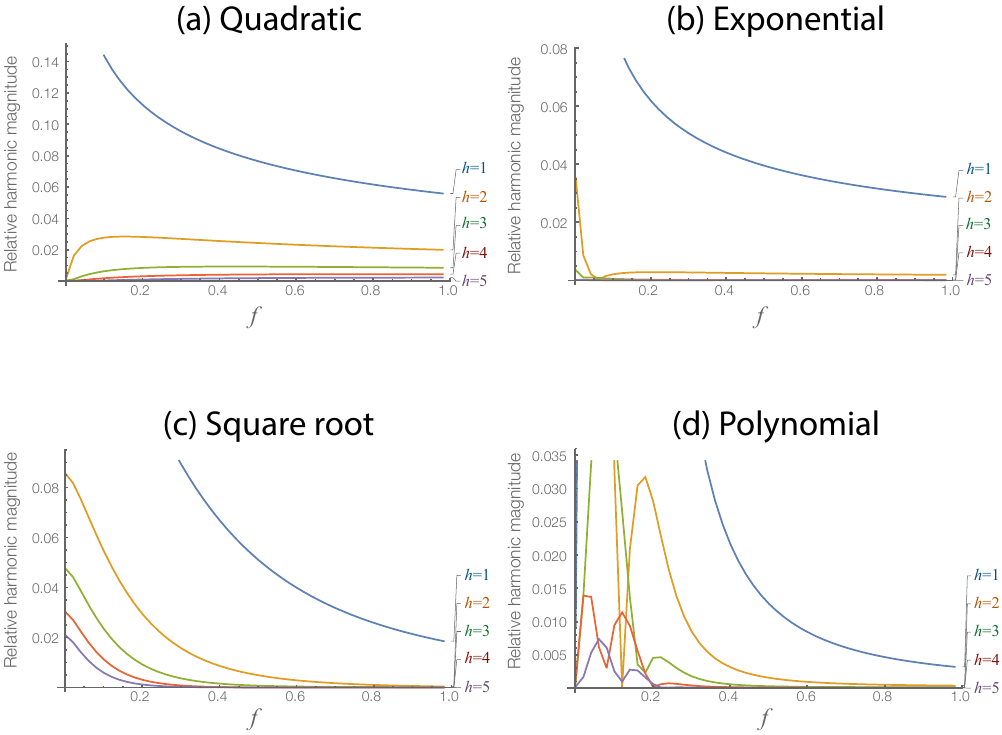} \\
  \caption{Relative changes in harmonic magnitude in terms of $f$ for the several considered non-linear transfer functions.}
  ~\label{f:feedback_harmonics}
  \end{center}
\end{figure*}

This interesting phenomenon was investigated further by performing symbolic quantification of the intensity of each of the first six harmonics with respect to the value of $f$.  Figure~\ref{f:feedback_harmonics} presents the respectively obtained results for each of the non-linear transfer functions considered above.  As expected, the feedback acts effectively in reducing the intensity of the first harmonic in all situations.  At the same time, several orders of harmonics are introduced in the case of the quadratic transfer function (Figure~\ref{f:feedback_harmonics}(a)), which only adds the first harmonic in the absence of feedback.  Interestingly, these added harmonics have a very slow decay.  In the case of the exponential (Figure~\ref{f:feedback_harmonics}(b)), the originally present harmonics are immediate and strongly reduced with $f$ (though their subsequent decays are, as in the case of the quadratic function, slow), corroborating the ability of feedback in linearizing this type of function.  Yet another behaviour is observed for the square root non-linearity (Figure~\ref{f:feedback_harmonics}(c)).   Now, the harmonics 2 to 5, which were originally present, decay in a way that is similar to that of harmonic 1 and consequently, are steadily reduced with the increase of $f$.  A surprisingly complex situation was obtained with respect to the polynomial transfer function, with the intensity of each harmonic progressing in widely different ways and at different scales of $f$.  All in all, the obtained results clearly indicate that completely different behaviours, regarding how the intensity of each harmonic varies with $f$, can be obtained for distinct types of non-linearities. In addition, these results also suggest that, when comparing the transfer function of different devices (e.g.~\cite{barbour1998cool}), it is interesting to refer also to no-feedback configurations. For instance, if a device with a perfect quadratic transfer function (which can only induce the first harmonic) is studied in presence of negative feedback, additional harmonics can be observed that are a consequence of the negative feedback action, not being a consequence of the intrinsic non-linear transfer function.

\subsection{Device parameter invariance}

In Section~\ref{s:negFeedback} and Appendix A, an equation associating the voltage gain $A_c^f$ with the circuit resistances and transistor parameters $\beta$ and $r_o$ was derived (Equation~\ref{eq:AcfApproxMainText}). This allowed the respective calculation of the sensitivity of the voltage gain with respect to variations of parameters $\beta$ (Equation~\ref{eq:SAcfVsBetaAppendix}) and $r_o$ (Equation~\ref{eq:SAcfVsRoAppendix}). In this section we use these equations to explore how the voltage gain sensitivity changes for different experimental conditions. We reiterate that the reason for doing such analysis is twofold. First, in situations where the sensitivity is low, we expect that the circuit will become insensitive to device variations. Second, parameters $\beta$ and $r_o$ can vary for a given device, depending on certain factors such as load line choice or temperature. Such a variation can, in turn, influence the linearity of the transfer function.

\begin{figure*}[]
  \begin{center}
  \includegraphics[width=0.8\linewidth]{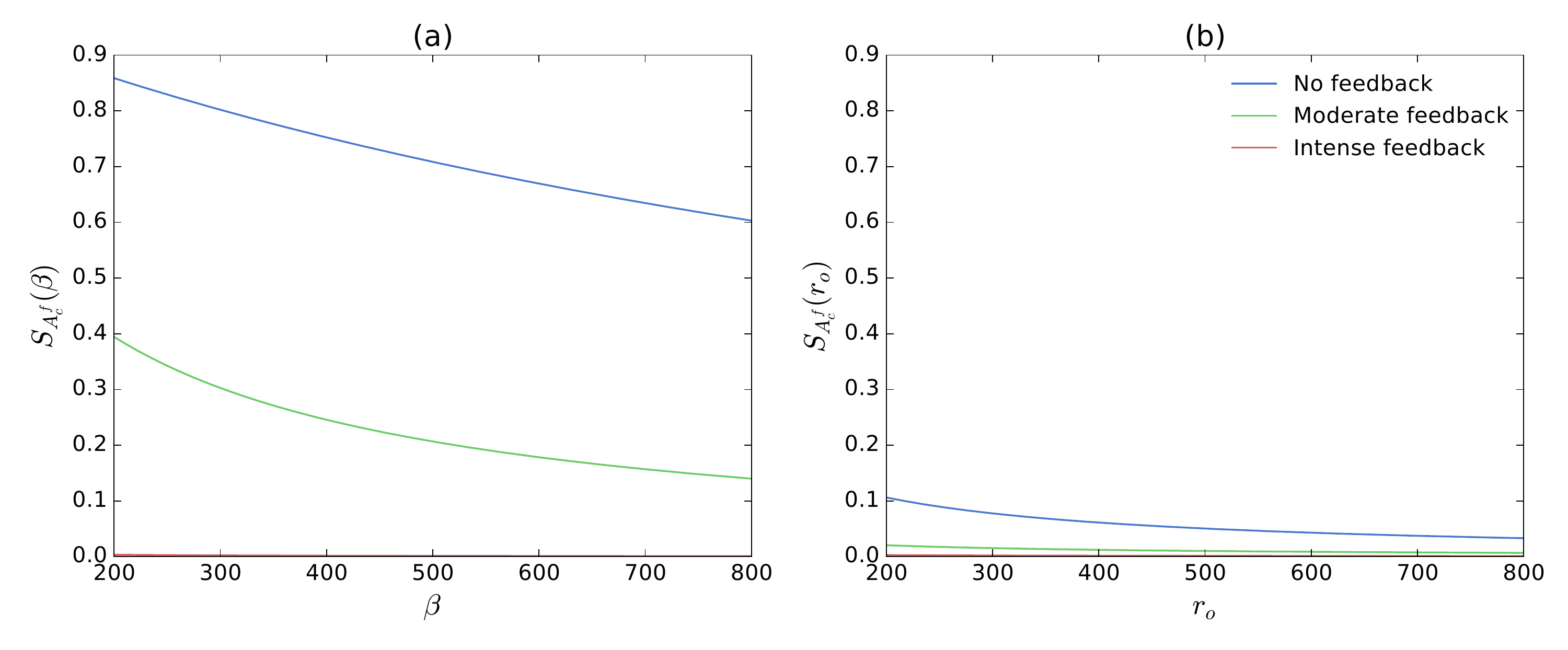} \\
  \caption{Sensitivity of current gain with respect to (a) $\beta$ and (b) $r_o$ for the three considered experiments.}
  ~\label{f:SAcf_vs_beta_ro}
  \end{center}
\end{figure*}

We start by considering the resistance values used in the experimental settings, which will be discussed in more detail in the next section, in the equations and verifying the influence of $\beta$ and $r_o$ on the sensitivity. Three situations were considered, differing according to the amount of feedback contained in the circuit: i) no feedback, ii) moderate feedback and iii) intense feedback. The resistances used in each situation are shown in Table~\ref{t:expParValues}. In Figure~\ref{f:SAcf_vs_beta_ro}(a) and (b) we show the sensitivity of $A_c^f$ with respect to, respectively, $\beta$ ($S_{A_c^f}(\beta)$) and $r_o$ ($S_{A_c^f}(r_o)$) for the three feedback levels. In the absence of feedback, $A_c^f$ is highly sensitive to parameter $\beta$, having a sensitivity of $S_{A_c^f}(\beta)\approx 0.7$. In addition, the sensitivity itself changes for different values of $\beta$. $S_{A_c^f}(\beta)$ decreases for larger values of the current gain, which indicates that a large $\beta$ is useful not only for additional amplification, but also for increased invariance of device parameter. For moderate feedback, the sensitivity becomes much smaller, specially for large $\beta$. In the intense feedback situation, parameter $\beta$ has no influence in the voltage gain. Regarding the sensitivity of $A_c^f$ with respect to $r_o$, a similar trend is observed. But in this case the sensitivity, even in the absence of feedback, is fairly low. Only very small values of $r_o$ would significantly influence the current gain in the considered experimental conditions.

\begin{figure*}[]
  \begin{center}
  \includegraphics[width=\linewidth]{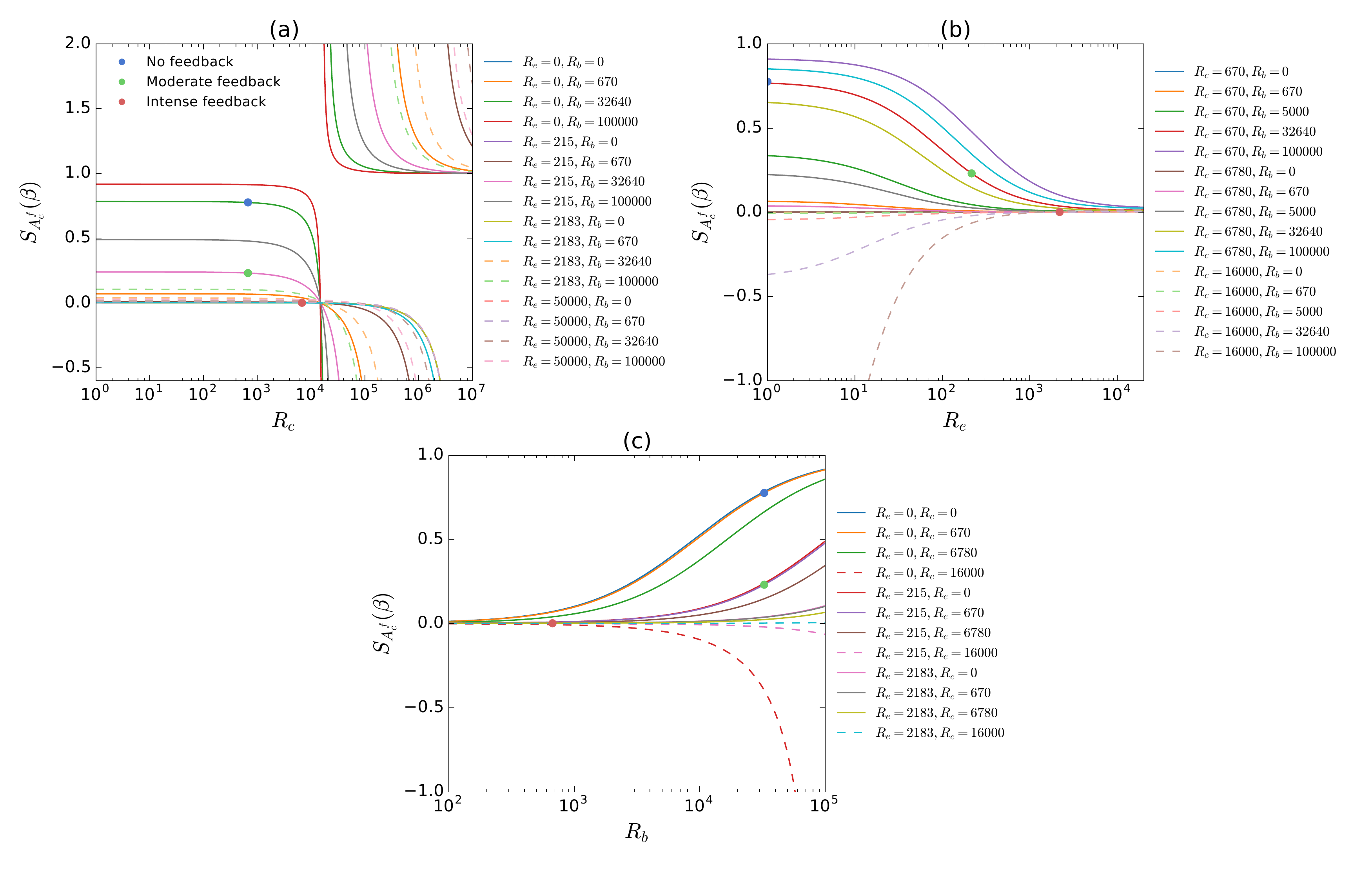} \\
  \caption{Voltage gain sensitivity with respect to $\beta$ considering each of the three involved resistances, i.e. (a) $R_c$, (b) $R_e$, and (c) $R_b$.  These results considered the experimentally obtained mean values $\beta=450$ and $r_o=14822\,\Omega$, as indicated in Table~\ref{t:parameterValuesExperiments}. All resistance values are given in Ohms.}
  ~\label{f:SAcf_beta_vs_Rc_Re_Rb}
  \end{center}
\end{figure*}

We also verify the change of sensitivity for distinct resistance values of the circuit. The results are shown in Figure~\ref{f:SAcf_beta_vs_Rc_Re_Rb}. In Figure~\ref{f:SAcf_beta_vs_Rc_Re_Rb}(a) we show the relationship between $S_{A_c^f}(\beta)$ and $R_c$. Each curve is respective to different combinations of $R_e$ and $R_b$. Markers indicate the parameters used in the experimental procedure. The plot indicates that small values of $R_c$ have no influence on the sensitivity. On the other hand, at a certain value of $R_c$, which depends on $R_b$ and $R_e$, a sharp divergence in sensitivity is observed. This happens because the denominator of Equation~\ref{eq:SAcfVsBetaAppendix} diverges when $R_c$ is 

\begin{equation}
R_c=r_o\frac{R_b+\beta(R_e+r_e)}{R_b}.
\end{equation}

Figure~\ref{f:SAcf_beta_vs_Rc_Re_Rb}(b) shows the dependence of sensitivity with $R_e$. For large $R_e$ values, as expected, $A_c^f$ becomes independent of $\beta$. For $R_e<1000\Omega$, $S_{A_c^f}(\beta)$ is influenced by $R_c$ and $R_b$. When $R_c>R_e+r_o+r_e$, the sensitivity becomes negative. In Figure~\ref{f:SAcf_beta_vs_Rc_Re_Rb}(c) we show $S_{A_c^f}(\beta)$ against $R_b$. In contrast to what was observed for resistance $R_e$, lower values of $R_b$ make the voltage gain nearly invariant to $\beta$. The sensitivity becomes surprisingly dependent on $R_b$ for larger values of this resistance.

\section{Experimental approach}
\label{s:experimentalApproach}

This section reports the experiments performed with real BJTs with respect to linearity and device invariance induced by negative feedback.  Several interesting results are described and discussed.

\subsection{Chosen devices}

Given the extremely large number of small signal transistors types available,
it is unavoidable to select a smaller group for our analysis.  A total
of 15 models of devices was used, incorporating some of the most popular
transistors.  Table~\ref{t:selectedTransistors} shows the characteristics, according to
datasheets, of the chosen transistor models (it should be observed that
some small variations can be found in specific datasheets).

\begin{table*}[]
\centering
\begin{tabular}{|c|cccccc|p{8cm}|}
\toprule
Name	& $V_{cbo}$(V)	& $I_{c,max}$ (mA)	& $P_D$(mW)	& $h_{fe}^{\min}$	& $h_{fe}^{\max}$ & $f_T$(MHz) & Description \\
\hline
\#01	& 60		& 800	& 500	& 30	& 300	& 300	& Switching, Linear \\
\#02	& 60		& 200	& 600	& 30	& 300	& 250	& Small load switching with high gain\\
\#03	& 60		& 600	& 600	& 20	& 300	& 250	& Switching and medium power amplification  \\
\#04	& 30		& 50	& 600	& 350	& 1400	& 50	& Low noise, high gain, general purpose \\
\#05	& 30		& 50	& 600	& 450	& 1800	& 50	& Low noise, high gain, general purpose \\
\#06	& 150		& 600	& 600	& 80	& 250	& 100	& General purpose, high-voltage\\
\#07	& 30		& 200	& 600	& 100	& 800	& 150	& General Purpose, low noise with good linearity  \\
\#08	& 30		& 200	& 600	& 200	& 800	& 150	& General Purpose, low noise with good linearity \\
\#09	& 80		& 100	& 500	& 100	& 800	& 300	& High-voltage, low noise   \\
\#10	& 50		& 100	& 500	& 100	& 800	& 300	& High-voltage, low noise  \\
\#11	& 30		& 100	& 500	& 100	& 800	& 300	& High-voltage, low noise  \\
\#12	& 30		& 100	& 500	& 100	& 800	& 300	& High-voltage, low noise  \\
\#13	& 50		& 100	& 500	& 100	& 800	& 300	& High-voltage, low noise  \\
\#14	& 60		& 1000	& 1000	& 50	& 150	& 100	& Switching and amplifier \\
\#15	& 50		& 100	& 600	& 400	& 1500	& 100	& Low noise, high gain \\
\hline
\end{tabular}
\caption{List of the chosen devices and respective features as obtained from datasheets, which include: $V_{cbo}$, the maximum voltage between collector and base; $I_{c,max}$, the maximum collector current; $P_D$, the maximum dissipation power; $h_{fe}^{\min}$ and $h_{fe}^{\max}$, respectively the minimum and maximum values of the current gain $h_{fe}$; and $f_T$, the current gain bandwidth product.}
\label{t:selectedTransistors}
\end{table*}

Two groups of devices --- corresponding to types \#07 and \#08, and to types \#09--\#13 --- belong to families of closely related transistors, varying mainly with respect to the collector-emitter isolation voltage.  Three
devices with higher $h_{fe}$ were also chosen, corresponding to cases
 \#04, \#05 and \#15.  A higher power transistor ($P_D$ = 1000mW), namely \#14, 
was also included, for generality's sake.  Model \#06 has a
higher isolation ($V_{cbo}$ = 150V.).  Two of the models, \#07 and \#08, have metal can 
package.  Typical end uses of the chosen devices, listed in the last
column of Table~\ref{t:selectedTransistors}, include switching and linear applications, low
noise, and general applications, corroborating the diversity of
finalities among the chosen small signal transistors.

All in all, a relatively representative selection including some uniformity
(among families) and diversity was adopted.  Four devices were
randomly chosen from respective lots. Therefore, a total of 60 transistors 
were analyzed.

\subsection{Experimental settings}

The adopted transistors were experimentally analysed using a microcontrolled
data acquisition system.  Three processing modules were used, the host PC,
a wifi Linux station, and a battery-powered, microcontroller module responsible for 4-channel DA and AD conversion.  More specifically, the former is responsible for applying voltages to the circuit, while the respective voltage outputs ($V_b$, $V_c$, $V_{bb}$ and $V_{cc}$) are synchronously latched by respective sample-and-holds and AD converted.  

The circuit used in the experiments is shown in Figure~\ref{f:amplifierWithFeedback}. 
The main parameters of this circuit correspond to the resistances $R_c$, $R_b$ and $R_e$. 
Three distinct situations were considered, differing according to the level of feedback 
contained in the circuit. These situations were: i) no feedback, ii) moderate feedback and iii) intense feedback. Recall that the expected level of feedback in the circuit is 
indicated by Equation~\ref{eq:feedbackApproxMainText}. Therefore, the three feedback levels were achieved by properly setting resistances $R_c$, $R_b$ and $R_e$ according to this equation. The resistances used for each case are indicated in Table~\ref{t:expParValues}. The table also shows the respective values of open loop gain $A_c$, feedback gain $f$ and feedback amount $fA_c$. In Table~\ref{t:pcaFeatures} we show the main properties of the device and the circuit measured in the experiments.  We developed a numerical procedure to estimate the BJTs parameters, which is described in Appendix B.

\begin{table*}[]
\centering
\begin{tabular}{p{50pt}p{60pt}p{88pt}p{70pt}}
\toprule
Parameter & No feedback & Moderate feedback & Intense feedback \\
\hline
$R_b$ & $32640\,\Omega$ & $32640\,\Omega$ & $670\,\Omega$ \\
$R_c$ & $670\,\Omega$ & $670\,\Omega$ & $6780\,\Omega$ \\
$R_e$ & $0\,\Omega$ & $215.8\,\Omega$ & $2183\,\Omega$ \\
$A_c$ & 7.7 & 7.7 & 332.2 \\
$f$ & 0.0 & 0.325 & 0.322 \\
$fA_c$ & 0.0 & 2.5 & 107.0 \\
\hline
\end{tabular}
\caption{Values of circuit resistances, open loop gain and feedback levels in the three feedback  situations considered in the experiments.}
\label{t:expParValues}
\end{table*}

\begin{table}[]
\centering
\begin{tabular}{lM{4.cm}l}
\toprule
Symbol & Name & Definition \\
\hline
$\beta$ & Current gain & Equation~\ref{eq:beta} \\
$A_v$ & Voltage gain & Equation~\ref{eq:AcfDefMainText} \\
$R_o$ & Output resistance & Equation~\ref{eq:ro} \\
$\text{THD}$ & Total harmonic distortion & Equation~\ref{eq:THDDef} \\
$V_a$ & Early voltage & Appendix B \\
\hline
\end{tabular}
\caption{Relevant device and circuit properties considered in the discussion of the experimental analysis.}
\label{t:pcaFeatures}
\end{table}

\begin{figure*}[]
  \begin{center}
  \includegraphics[width=0.85\linewidth]{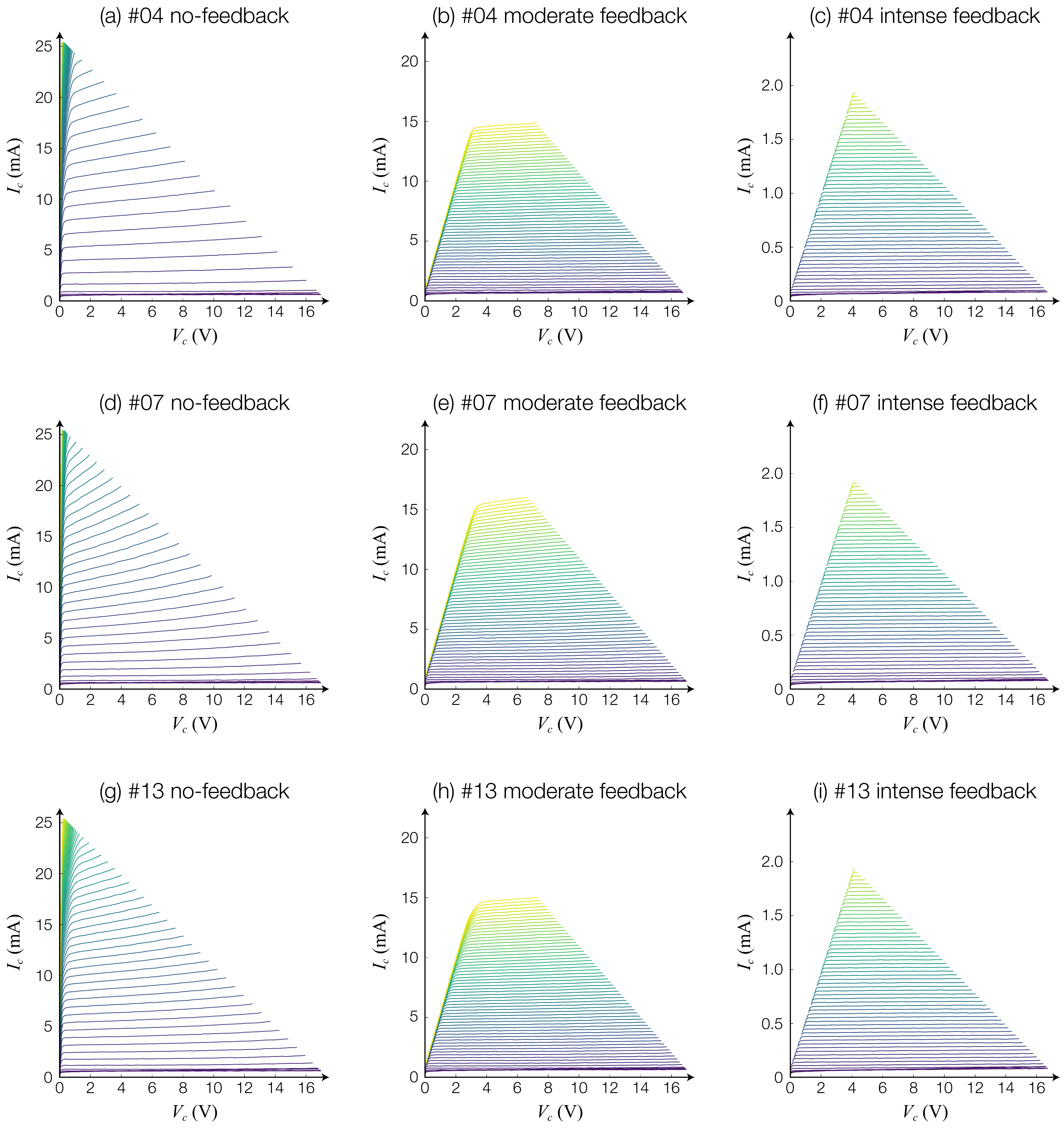} \\
  \caption{Illustration of the experimentally obtained isolines in the $I_c \times V_c$ space with respect to samples from the \#04, \#07 and \#13 transistors types.  The columns correspond, respectively, to no-feedback, moderate feedback, and intense feedback.  See the text for discussion. In the case of intense feedback, the portions of the isolines along the saturation region were left out for clarity's sake.}
  ~\label{f:isolinesExample}
  \end{center}
\end{figure*}

Figure~\ref{f:isolinesExample} illustrates the type of data obtained in our experiments.   More specifically, the space $I_c \times V_c$, with respective isolines, are presented for a sample from each of the transistors types \#04, \#07, and \#13 in the experiments using no-feedback (Figures~\ref{f:isolinesExample}(a), (d) and (g)), moderate feedback (Figures~\ref{f:isolinesExample}(b), (e) and (h)), and intense feedback (Figures~\ref{f:isolinesExample}(c), (f) and (i)).  It is clear from an analysis of the first column of this figure that the original devices present great variation of properties in absence of feedback, as reflected by the straightness and slope of the isolines (corresponding to the inverse of $R_o$), the different saturation regions, and separation between the isolines (related to the current gain).  The addition of moderate feedback, illustrated in the middle column of Figure~\ref{f:isolinesExample}, promoted an impressive normalization of the isolines, which now have more similar straighness, slope, and separation, as well as more standardized saturation regions.  At the same time, observe that $R_o$ increased substantially with the addition of moderate feedback.  The intensification of negative feedback, shown in the third column of Figure~\ref{f:isolinesExample}, clearly yielded an even more impressive standardization of the isolines, at the cost of reduced gain and very high $R_o$.  It is also interesting to observe the effect of negative feedback in making the isolines more parallel, therefore substantially decreasing the Early voltage.

\subsection{Measurement by measurement analysis}
\label{s:singleParameterInvariance}

\begin{figure*}[]
  \begin{center}
  \includegraphics[width=\linewidth]{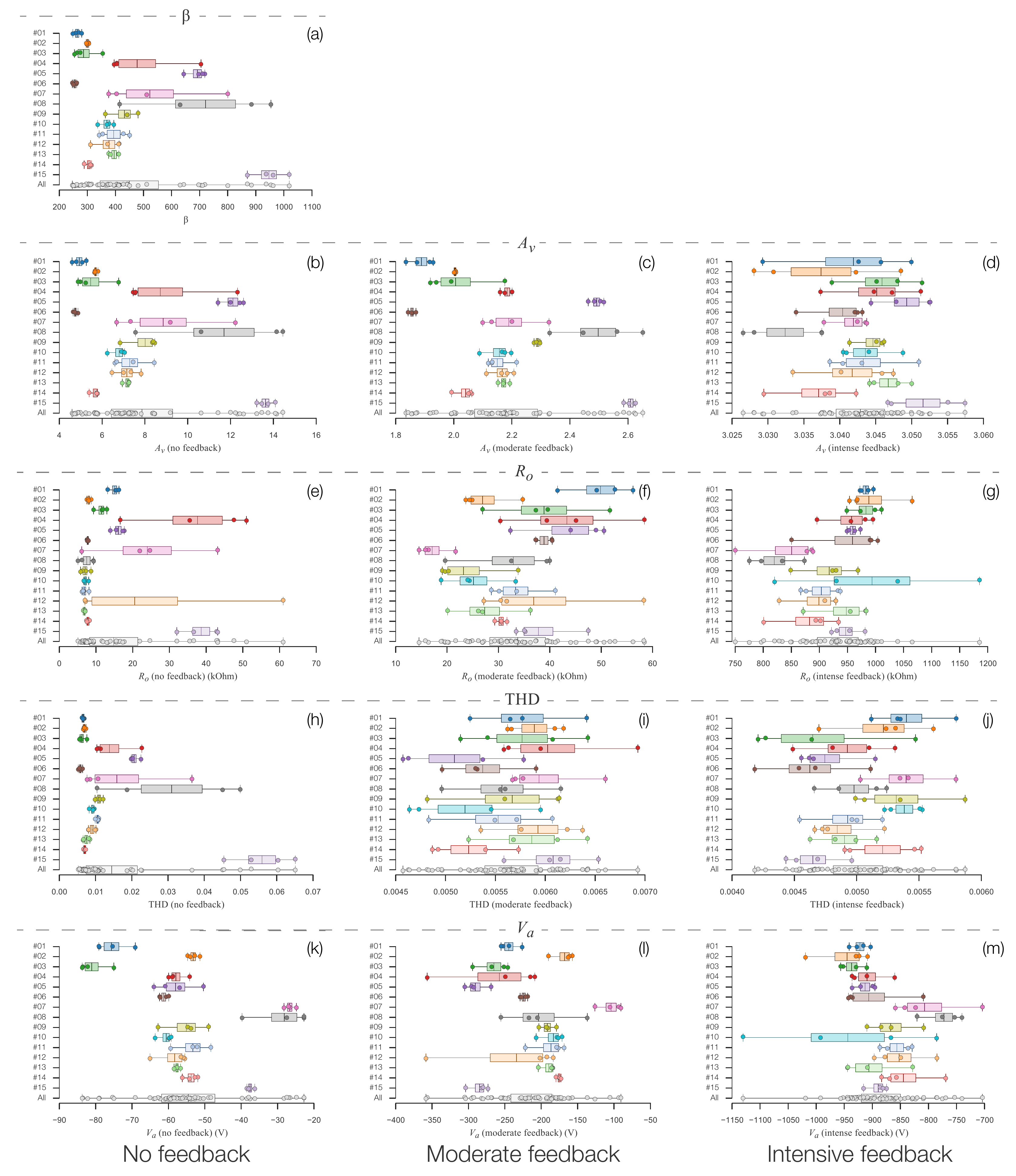} \\
  \caption{Boxplots indicating the experimentally obtained distributions of device and circuit properties for each family of transistor. The first, second and third columns of plots are relative to, respectively, the absent, moderate and intense feedback conditions.}
  ~\label{f:boxplots}
  \end{center}
\end{figure*}

We start the experimental analysis of the transistors by taking into account their average, standard deviation and coefficient of variation values, which are illustrated in Figures~\ref{f:boxplots}(a)-(m).  The scatter distances between the transistor types for each measurement were also obtained and are presented in Table~\ref{t:scatterDistances}.  A general analysis of these results readily indicates that, in most cases, the variance of the properties for each transistor model tends to be small, implying in respectively compact groups and parameter coherence.  This is the case, for instance, of the values of $\beta$, shown in Figure~\ref{f:boxplots}(a), which are respective only to the experiment without feedback (recall that $\beta$ is a constant of the device and does not depend on the circuit configuration).  With exception of transistors \#04, \#07 and \#08, all other groups are compact, indicating that each of those models have typical, characteristic values of $\beta$.  This is reflected in the respective relatively large scatter distance (Table~\ref{t:scatterDistances}) value of 5.3 obtained for this case.

\begin{table*}[]
\centering
\begin{tabular}{cccccc}
\toprule
 Feedback intensity & $\beta$ & $A_v$ & $R_o$ & THD & $V_a$ \\
\hline
Absent & 5.3 & 5.4 & 1.8 & 4.8 & 17.3 \\
Moderate & - & 16 & 1.6 & 0.49 & 2.8  \\
Intense & - & 0.94 & 0.94 & 0.94 & 1.0 \\
\hline
\end{tabular}
\caption{Scatter distances for each of the considered measurements with respect to the three experiments.}
\label{t:scatterDistances}
\end{table*}

The above results contradict the common idea that transistors have widely varying parameters, especially $\beta$. However, it should be always borne in mind that the results in this article are, in principle, specific and restricted to considered devices, which originate from a same specific lot.  The average $\beta$ of each model has good agreement with the maximum gain $h_{fe}^{\max}$ specified in respective datasheets (see a compilation of these values in Table~\ref{t:selectedTransistors}). We verified that the Pearson correlation between these two quantities is 0.79, which corroborates this result.    It also follows from Figure~\ref{f:boxplots}(a) that most transistors have values of $\beta$ comprised between 200 and 500, which is in general agreement with common experimental practice.  Some groups, such as the \#02, and \#06 present surprisingly consistent values of $\beta$.  Such a coherence is also manifested in the overlap between the distributions of $\beta$ obtained for the family of transistors \{\#09, \#10, \#11, \#12, \#13\}.  

The measurements presented in the other figures (i.e. Figures~\ref{f:boxplots}(b)-(m)) are respective to each of the three performed experiments (i.e. without feedback, moderate feedback and intense feedback).  Figure~\ref{f:boxplots}(b) shows the averages and standard deviations of voltage gain in the no-feedback situation.  The mean, standard deviation and coefficient of variation of the considered BJT properties are given in Table~\ref{t:parameterValuesExperiments}.  These results are similar to those obtained for $\beta$, with most groups exhibiting a relatively compact distribution of values and mean values distributed in the interval from 4 to 16.  The average gain when considering all transistors is $\mu_{A_v}=7.9$, and the standard deviation is $\sigma_{A_v}=2.9$. The respective coefficient of variation of the current gain is $CV_{A_v}=0.37$.   This result is consistent with the scatter distance of 5.4 obtained for this measurement.  The incorporation of moderate feedback (Figure~\ref{f:boxplots}(c)) reduced the variation of $A_v$ to a smaller interval, ranging from 1.8 to 2.8, resulting in a coefficient of variation of $CV_{A_v}=0.1$.  However, the distance between groups, as quantified by the respective scatter distance, increased to 16.   For the intense feedback situation presented in Figure~\ref{f:boxplots}(d), the average, standard deviation and coefficient of variation become, respectively, $\mu_{A_v}=3.0$, $\sigma_{A_v}=0.01$ and $CV_{A_v}=0.002$. The average value is in good agreement with the theoretically expected value of $A_c^{f}=3.1$ obtained by using Equation~\ref{eq:AcfApproxMainText}.  In this case, negative feedback was effective in considerably reducing the scatter distance to 0.94.  The values of $R_o$ obtained in the three types of experiments are shown in Figures~\ref{f:boxplots}(e)-(g).  In the absence of feedback, most of the transistor models have similar $R_o$ values of about $10000\Omega$, which is identical to the transistors constant $r_o$. For this experiment, we obtained $CV_{R_o}=0.90$ and scatter distance of 1.8. The incorporation of moderate feedback (Figure~\ref{f:boxplots}(f)) not only uniformized the values of $R_o$, resulting in $CV_{R_o}=0.32$, but also increased them by a factor of 3 or 4.  The scatter distance was slightly reduced to 1.6.  More intense normalization was achieved for intense feedback (Figure~\ref{f:boxplots}(g)), where the coefficient of variation becomes $CV_{R_o}=0.08$ and the scatter distance decreases to 0.94, but this was achieved at the expense of a substantial increase of $R_o$ to values of nearly $1M\Omega$.  

Interesting results were obtained also regarding THD values.  Without feedback, the THD values are, on average, $\mu_{\mathrm{THD}}=0.014$, i.e. $1.4\%$.  Particularly high THD values were obtained for the transistors \#15, which are designed for high gain applications. Also, due to some transistors presenting high THD values, a high coefficient of variation, $CV_{\mathrm{THD}}=0.986$, was obtained.  The scatter distance for this measurement is relatively high, being equal to 4.8.  With application of moderate feedback, the THDs were substantially uniformized ($CV_{\mathrm{THD}}=0.095$ and scatter distance equal to 0.49), but the THDs values were reduced to only about half of their original values, including the \#15 case.  The application of intense feedback did not contribute to noticeable additional reduction of the THD values ($CV_{\mathrm{THD}}=0.08$ and scatter distance equal to 0.94).  The obtained model average and standard deviation values of Early voltages are shown in Figures~\ref{f:boxplots}(k)-(m).  A great variation of $V_a$ values was obtained, ranging from $-90V$ to $-20V$ ($CV_{V_a}=-0.27$) around an average value of $\mu_{V_a}=-54.6V$.  At the same time, the variation inside each group is noticeably compact and similar among the several groups of devices.  This implied the largest scatter distance among all measurements, being equal to 17.3.  The incorporation of moderate feedback increased substantially the values of $V_a$, with overall average $\mu_{V_a}=-213V$. Surprisingly, the coefficient of variation remained the same as in the case of no feedback, but the separation between the groups was substantially reduced as indicated by the respective scatter distance of 2.8.   Intense feedback had a remarkable effect in normalizing $V_a$ to values around $-900V$, resulting in compact groups except for \#10, as reflected in the respective scatter distance of 1.0. The coefficient of variation for this case was $CV_{V_a}=-0.08$.

\begin{table*}[]
\centering
\begin{tabular}{lrrr}
\toprule
Property    &  Mean     &  Std. dev. &  Coef. var.\\
\hline
$\beta$        &  449      &  210      &  0.467     \\
$A_v$ (nF)  &  7.86     &  2.89     &  0.367     \\
$A_v$ (mF)  &  2.18     &  0.219    &  0.100  \\
$A_v$ (iF)  &  3.04     &  0.00682  &  0.00224   \\
$r_o$       &  $14822\,\Omega$    &  $13270\,\Omega$    &  $0.895\,\Omega$     \\
$R_o$ (mF)  &  $33731\,\Omega$    &  $10939\,\Omega$    &  $0.324\,\Omega$     \\
$R_o$ (iF)  &  $932774\,\Omega$   &  $73416\,\Omega$    &  $0.0787\,\Omega$    \\
THD (nF)    &  0.0145   &  0.0143   &  0.986     \\
THD (mF)    &  0.00566  &  0.000540 &  0.0953    \\
THD (iF)    &  0.00501  &  0.000401 &  0.0800    \\
$V_a$ (nF)  &  $-54.55\,V$   &  $14.8\,V$     &  $-0.271\,V$    \\
$V_a$ (mF)  &  $-213.45\,V$  &  $57.0\,V$     &  $-0.267\,V$    \\
$V_a$ (iF)  &  $-884.78\,V$  &  $70.2\,V$     &  $-0.0794\,V$   \\
\hline
\end{tabular}
\caption{Mean, standard deviation and coefficient of variation of device and circuit properties obtained in the experimental analysis. nF, mF and iF refer to, respectively, absent, moderate and intense feedback.}
\label{t:parameterValuesExperiments}
\end{table*}

The individual measurement analysis, especially the respective scatter distances in Table~\ref{t:scatterDistances}, provide principled ground for discussing the overall effect of negative feedback on the reduction of device parameter dependence.  More specifically, as already discussed, we have that a substantial reduction of scatter distances were obtained in most cases.  However, the efficiency of feedback in reducing such a dependence was not uniform regarding every measurement.  For instance, the group separation (and therefore device dependence) defined by the Early voltage was quickly reduced from 17.3 to 2.8 with the increase of negative feedback.  On the other hand, the group separation related to $A_v$ increased substantially with application of moderate feedback.  As it is clearly confirmed by visual inspection of Figure~\ref{f:boxplots}(i), where the groups are almost indistinguishable, the greatest reduction of the dependence to device parameters was obtained with respect to the THD in the case of moderate feedback.  Though low scatter distance values were obtained for all measurements in the intense feedback case, it can be observed from Figure~\ref{f:boxplots} that several transistor types remained segregated, indicating that even intense feedback is unable to completely erase the memory with respect to most measurements.

\subsection{Parameter invariance}
\label{s:multipleParameterInvariance}

\begin{figure*}[]
  \begin{center}
  \includegraphics[width=\linewidth]{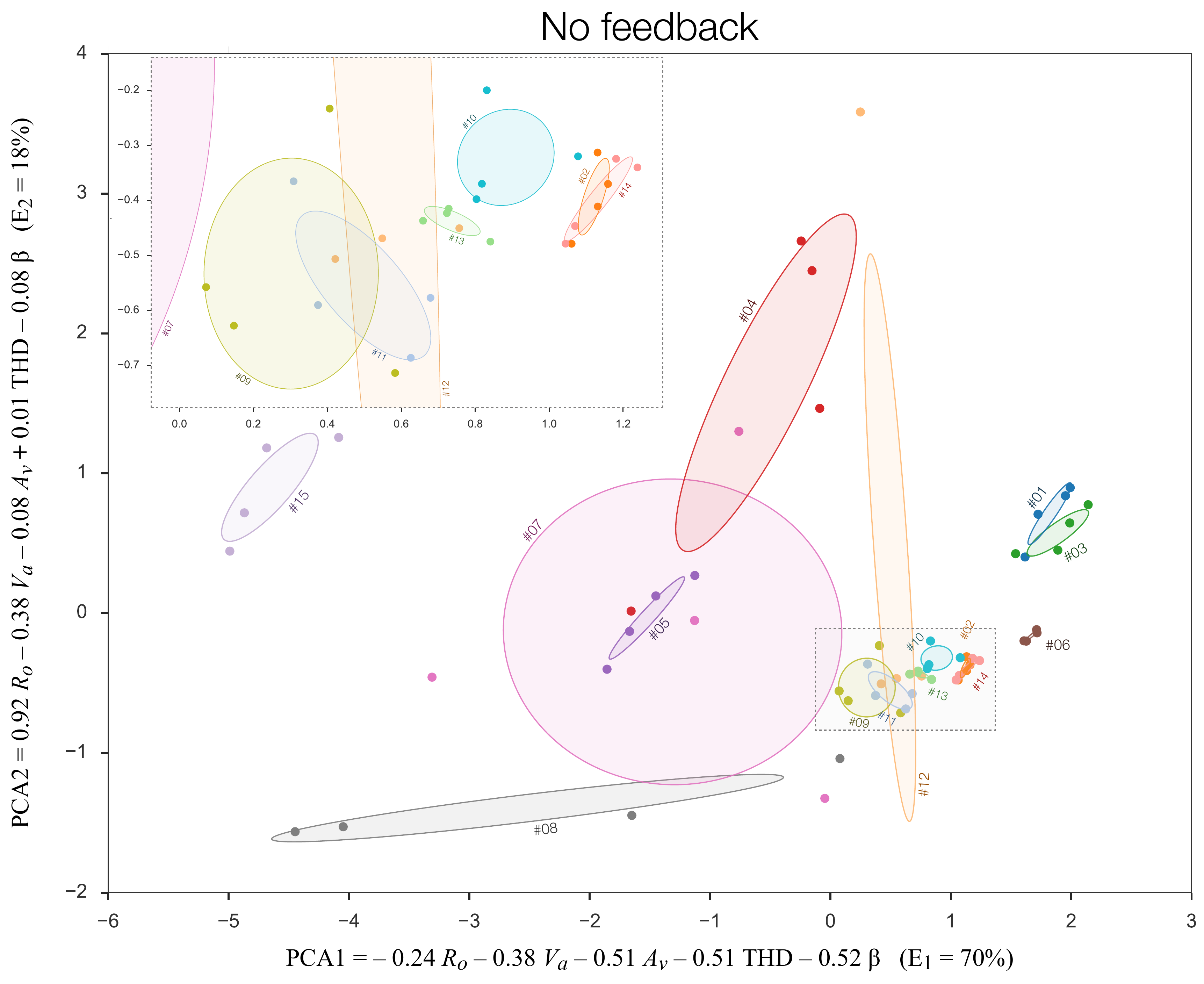} \\
  \caption{PCA of the $\beta$, $A_v$, $R_o$, THD and $V_a$ circuit properties measured without feedback. Each point represents a transistor and points are colored according to the type of the transistor. Confidence ellipses are also shown for each transistor type. The information retained by each axis is indicated in parenthesis next to the axis name. The inset shows a zoomed-in version of the points inside the indicated grey dashed box.}
  ~\label{f:newPCA-nF-v2}
  \end{center}
\end{figure*}

\begin{figure*}[]
  \begin{center}
  \includegraphics[width=\linewidth]{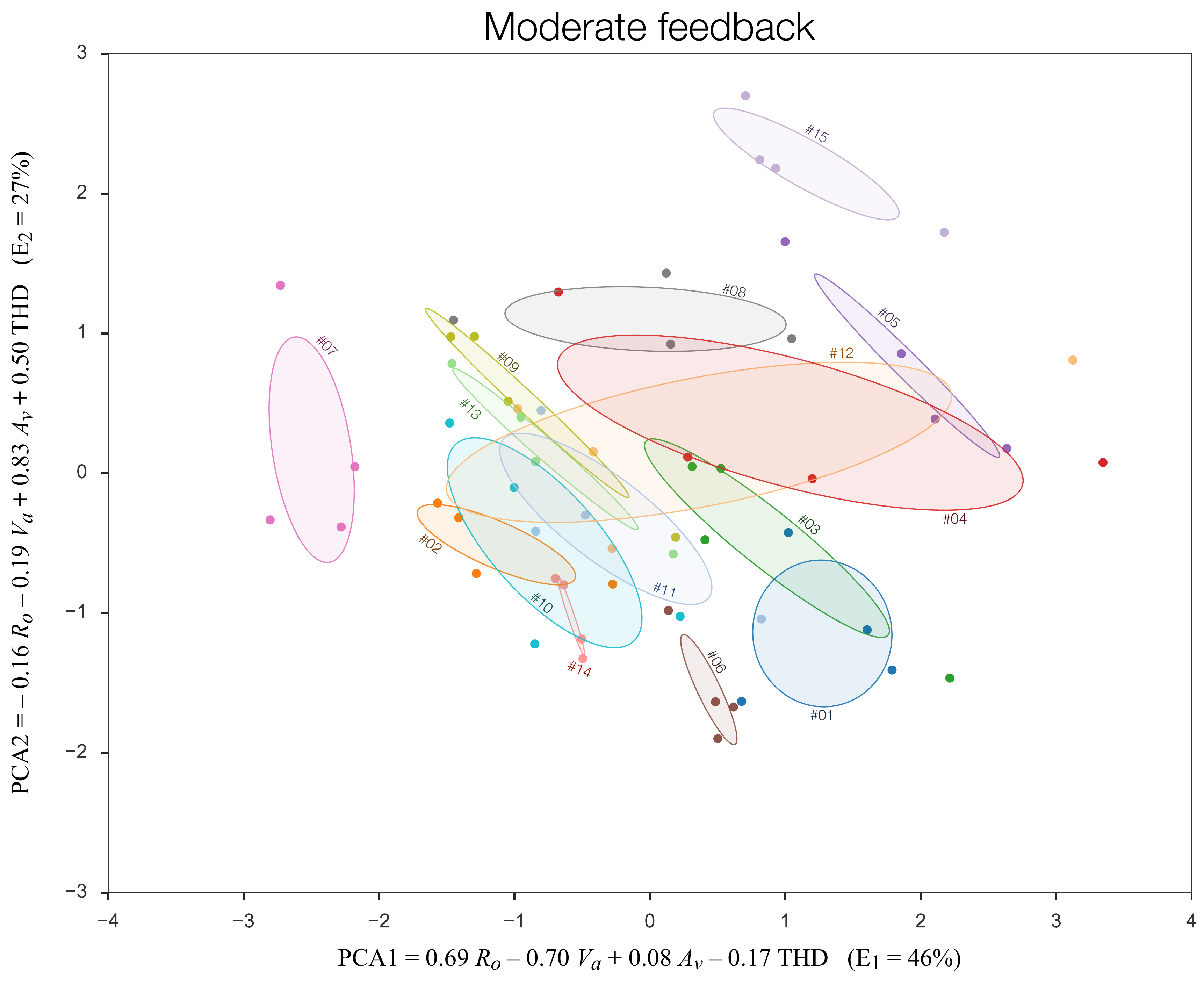} \\
  \caption{PCA of the $\beta$, $A_v$, $R_o$, THD and $V_a$ circuit properties measured with moderate feedback. Each point represents a transistor and points are colored according to the type of the transistor. Confidence ellipses are also shown for each transistor type. The information retained by each axis is indicated in parenthesis next to the axis name.}
  ~\label{f:newPCA-mF-v2}
  \end{center}
\end{figure*}

\begin{figure*}[]
  \begin{center}
  \includegraphics[width=\linewidth]{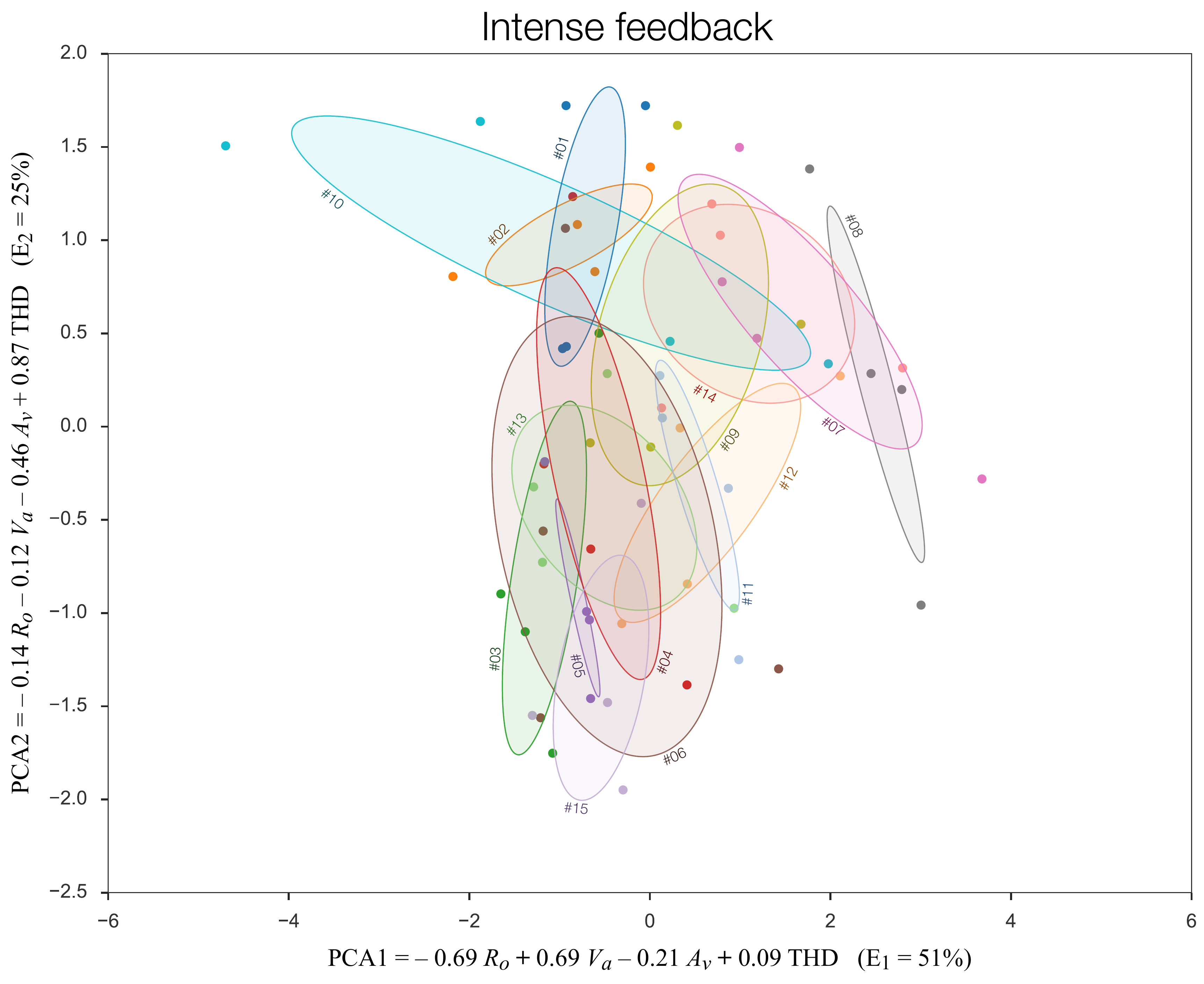} \\
  \caption{PCA of the $\beta$, $A_v$, $R_o$, THD and $V_a$ circuit properties measured with intense feedback. Each point represents a transistor and points are colored according to the type of device. Confidence ellipses are also shown for each transistor type. The information retained by each axis is indicated in parenthesis next to the axis name.}
  ~\label{f:newPCA-iF-v2}
  \end{center}
\end{figure*}

Figures~\ref{f:newPCA-nF-v2},~\ref{f:newPCA-mF-v2} and~\ref{f:newPCA-iF-v2} present the PCAs obtained from all measurements for all transistors in the three main experiments, namely without feedback, moderate feedback and intense feedback. Each group of devices, respective to a given type, are represented by different colors, and ellipses are used to highlight the position and extension of the respective variances.  

As is readily verified from Figure~\ref{f:newPCA-nF-v2}, the groups of transistors tend to clusterize strongly, indicating that their respective properties are characteristic and that, consequently, it makes sense to speak in terms of differentiated transistor types.  Moreover, this result clearly paves the way to the subsequent, more systematic quantitative investigation on how much feedback can act in promoting device invariance.  If the invariance is large, the points defined by the measurements of all devices should merge into a single cluster, indicating that the operation would be irrespective of the type or individual transistor.  It also follows from Figure~\ref{f:newPCA-nF-v2} that, despite the good separation between the groups, the properties of different types of transistors tend to present varying degrees of dispersion.  Remarkably, $88\%$ of the total variation of the measurements was captured by the first two principal components, indicating a high degree of correlation between the considered measurements.  The weights of the contributions of the measurements to the first axis are relatively high and have similar magnitude, indicating that this first axis takes into account all measurements in an almost equal manner.  The second principal axis is dominated by the values of $R_o$.

The effect of the introduction of moderate feedback is clearly inferred from Figure~\ref{f:newPCA-mF-v2}.  Though many of the groups of devices are still well-separated, the overall distance between them was clearly reduced, indicating enhanced invariance to device properties.  The respective scatter distances (now considering all measurements) provide a quantification of such an enhancement, varying from 35.94 to 24.31.  An additional effect of feedback was to uniformize the variation of properties also within each group, in the sense that the extent of the ellipses are more comparable one another than in the case without feedback.  A total of $73\%$ of the total variation of the measurement data is now accounted for by the first two principal axes.  The first component is dominated by $R_o$ and $V_a$, and the second by $A_v$ and $THD$.  It is interesting to observe that the nature of these two axes are therefore different from that obtained in the absence of feedback.  This shows that the uniformization implemented by feedback is not relatively uniform in all measurements.

Figure~\ref{f:newPCA-iF-v2} shows the PCA obtained in the case of intense feedback.  Strong overlap is now observed between almost every pair of groups, corroborating the ability of feedback in making the behavior of the circuit in question less dependent of the device parameters.  This is directly reflected in the reduction of the scatter distance value of 24.31 (in the case of moderate feedback) to only $3.73$.  However, such a remarkable invariance was obtained at the expense of extremely high output resistances (in the order of $1 M\Omega$), which strongly constrains the possible applications of this circuit configuration.  Now, $76\%$ of the original variance is explained by the first two axes.  As with moderate feedback, the first axes reflects $R_o$ and $V_a$, while the second axis is dominated by $A_v$ and $THD$.  It is interesting to observe that the composition of the two first axis changed substantially while moving from no-feedback to moderate feedback, which implied in a relatively small invariance enhancement, but remained similar between moderate and intense feedback while the variance decreased steeply.  This can be interpreted as being a consequence of the fact that the moderate feedback is enough to uniformize the several measurements to a level from which further variance reductions become more similar and therefore unable to change the orientation of the PCA projection.

\subsection{Correlation and graph analysis}

\begin{figure*}[]
  \begin{center}
  \includegraphics[width=\linewidth]{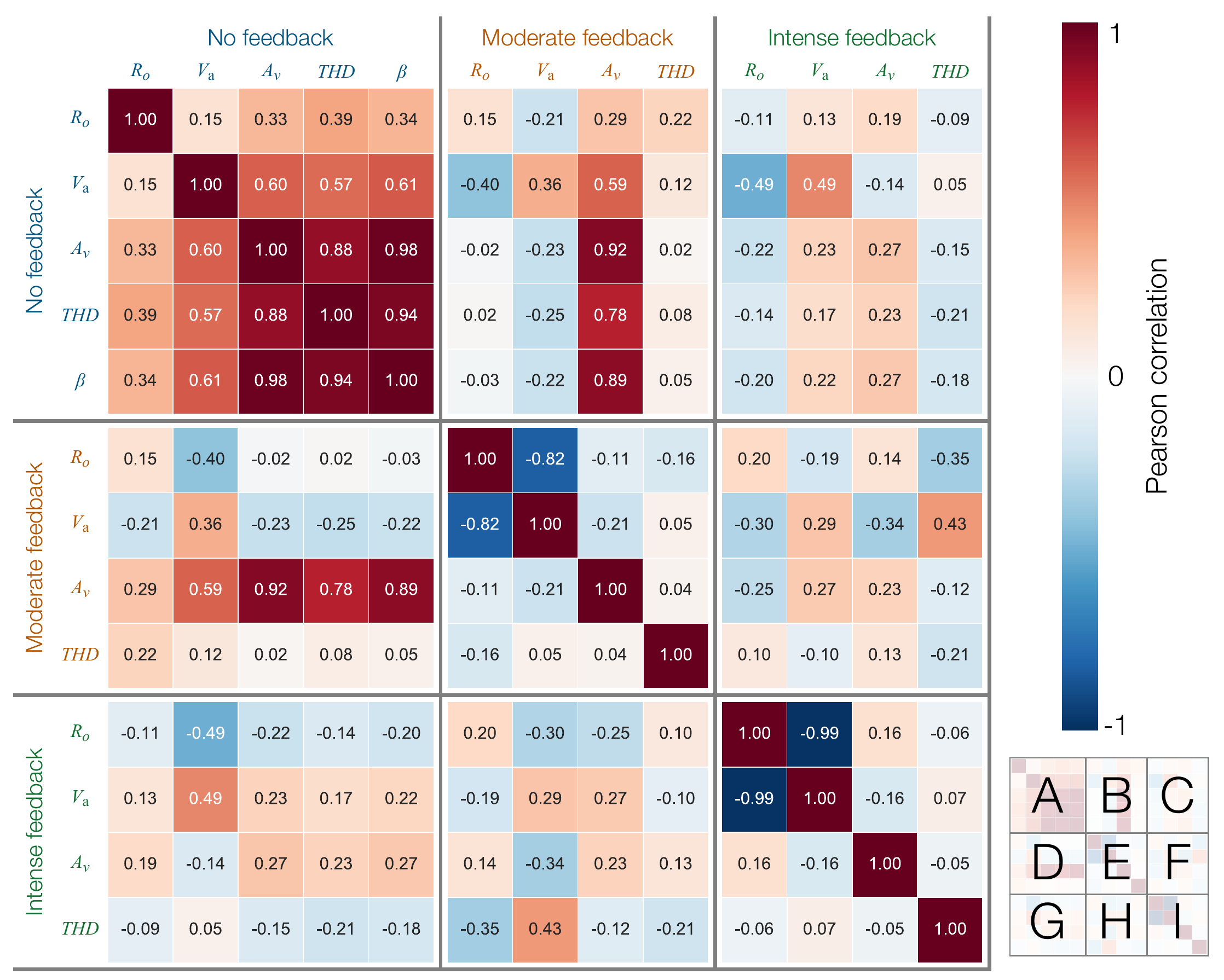} \\
  \caption{Pairwise Pearson correlation values between all device and circuit properties measured in the experiments. Each block of the matrix corresponds to a combination of feedback conditions (absent, moderate or intense). These blocks are named according to the legend shown in the bottom right corner of the figure.}
  ~\label{f:correlationMatrix-v2}
  \end{center}
\end{figure*}

So far, we presented and discussed the obtained BJT measurements in a single-variable (Section~\ref{s:singleParameterInvariance}) and multiple-variable fashion (Section~\ref{s:multipleParameterInvariance}).  In this section we complement that analysis by explicitly considering pairwise relationships between the measurements, as quantified by respective Pearson correlation coefficients. The results are summarized in Figure~\ref{f:correlationMatrix-v2}, which includes a bottom-right legend identifying its main blocks.  For simplicity's sake, the discussion of these results is performed first for each of the three types of experiments (i.e. blocks A, E, and I). The relationships between experiments (i.e. B, C, D, F, G and H blocks) are discussed subsequently.

In the case of the first experiment (block A), we have intense correlations for all cases, except the pairwise relations involving $R_o$.  A particularly strong relationship is observed between $\beta$ and THD, which indicates that higher current gains tend to undermine linearity.  The lowest correlation was obtained between $R_o$ and $A_v$.  The correlations obtained for the moderate feedback experiment are markedly different from the previous experiment, with low values throughout, except for the pair $R_o$ and $V_a$, which is now strongly negative.  This implies, for instance, that $A_v$ becomes less related to THD, implying in a less marked relationship between linearity and voltage gain.  The pairwise relationships obtained for the third experiment (intense feedback), presented in block I, are generally similar to the just discussed case (moderate feedback).  Interestingly, the correlation between $R_o$ and $V_a$ is very high in this experiment.  This can be understood as follows.  In the case of intense feedback, the reference isoline  (please refer to the constructions in Appendix B) tends to be the same in all cases.  At the same time, the isolines tend to be nearly parallel one another (i.e. they have the same $R_o$).  So, the values of $V_a$ will be mostly defined by $R_o$, hence the observed intense negative correlation.

The correlations between measurements obtained in different experiments are discussed as follows.  In block D (or B, as the correlations are symmetric), weak correlations are generally observed, with exception of the cases involving $A_v$.  An interesting way to interpret this result is by comparing with the no feedback experiment.  More specifically, in that case we had intense correlations in most cases, including $A_v$.  The addition of moderate feedback provided good invariance to most of those measurements, resulting in low correlations.  However, this promotion of invariance was less effective regarding $A_v$, implying in a memory effect regarding this measurement.  The other two cases (i.e. blocks G and H), we have low correlations between all pairs of measurements, indicating that intense feedback was capable of achieving, virtually, independence of device parameters.

\begin{figure}[]
  \begin{center}
  \includegraphics[width=\linewidth]{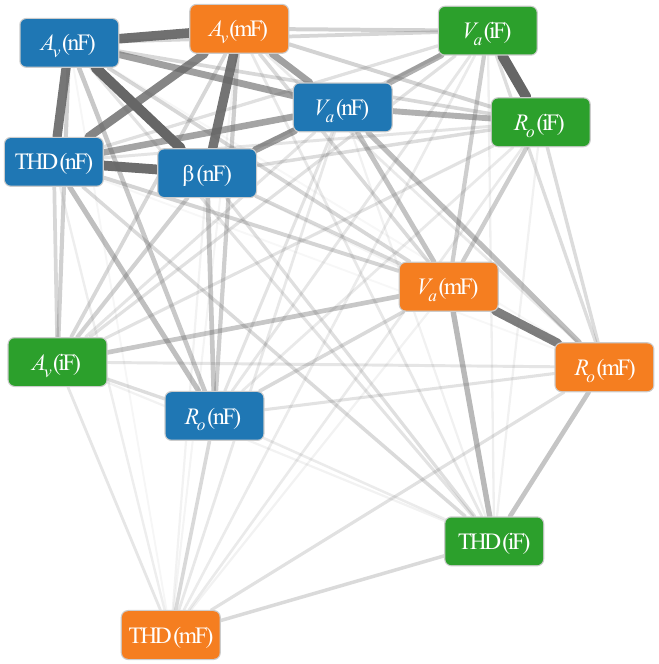} \\
  \caption{Correlation graph of the circuit and device properties considered in the experiments. Each node represents a property measured in the indicated feedback condition, and edges indicate the absolute Pearson correlation value between pairs of properties.}
  ~\label{f:meaGraph}
  \end{center}
\end{figure}

A summary of the results obtained in the experimental analysis is presented as a graph in Figure~\ref{f:meaGraph}. Each node of the graph corresponds to a property measured in the indicated feedback condition, namely: i) no feedback (nF, blue), ii) moderate feedback (mF, orange) and iii) intense feedback (iF, green). The width of the connection between a pair of nodes represents the absolute value of the Pearson correlation between the respective properties. This graph immediately reveals the presence of an intensely interconnected group of five properties, namely current gain ($\beta$), voltage gain without feedback ($A_v$~(nF)), THD without feedback ($THD$~(nF)), voltage gain with moderate feedback ($A_v$~(mF)), and Early voltage without feedback ($V_a$~(nF)). The strong association between voltage and current gain in the absence of feedback was already expected. The fact that THD and $A_v$~(mF) also posses such a strong relationship indicates that not only linearity is largely dependent on gain, but also that the current gain in the presence of moderate feedback is strongly related to properties measured in absence of feedback. A second group of properties can be observed in the graph. This group is composed of three properties, namely $V_a$~(nF), $V_a$~(iF) and $R_o$~(iF), with $V_a$~(nF) providing the interconnection between these two groups.  Interestingly, moderate feedback made THD largely uncorrelated with all other properties.  The obtained graph also incorporates the already discussed pairwise relationship between $V_a$~(mF) and $R_o$~(mF).

\section{Concluding remarks}

Negative feedback is a quintessential concept that permeates great part of natural and artificial systems.  It can be used to enhance both linearity and invariance to device parameters.  As such, negative feedback has been instrumental in paving the way to the widespread adoption of semiconductor transistors in modern electronics, which is especially critical given the large variability of the properties of such devices even when restricted to a specific transistor type.   So, the achievement of effective linear systems largely depends on the application of negative feedback.  Yet, at the same time, thousands of transistors types have become available commercially, even in the same specific application niche, motivating interesting questions.   First, if transistors are known to have broad variability of parameters even within the same type, would such groups be discernible as a whole?   Second, how effectively can negative feedback make a circuit independent of the original device parameters, and so produce a completely device-independent implementation?  Third, how is linearization affected as a circuit becomes more device invariant?  The current work addressed such questions by applying analytical and experimental approaches, and several interesting results have been obtained.

First, it was shown that the enhancement of linearity by negative feedback depends on the type of non-linearity in the original transfer functions.  For instance, it was illustrated that, in the case of a quadratic function, although feedback increases the overall linearity, it also adds new harmonics to the signal.  A different effect was observed for the exponential non-linearity, with all the harmonics induced by that type of transfer function being readily attenuated as $f$ is increased.  We also investigated, in theoretical fashion, how the sensitivity of several properties of a class A common emitter amplifier varies with the circuit configuration, and the results revealed a considerably complex situation regarding the interdependence of the three involved resistors ($R_c$, $R_e$, and $R_b$) on the resulting properties of the system, particularly voltage gain.  This corroborates the fact that linear design is a particularly challenging endeavour since the operation of the circuit will critically depend on the combined effect of the chosen components.  With this respect, we emphasise the potential of modern concepts and methods from multivariate statistics and artificial intelligence as a means to be tried for the optimised design of such circuits.

As a matter of fact, concepts from these areas, especially principal component analysis~\cite{joliffe1992principal}, pattern recognition~\cite{bishop2006pattern}, and complex networks~\cite{newman2010networks}, have been applied in the experimental section of this work as a means to investigate further the effects of feedback regarding parameter independence and linearization of transfer functions.  This has been done by first choosing a set of representative small signal transistors types, and inferring the respective parameters (individually or in a common emitter amplifier), characteristic and transfer functions by using a microcontrolled experimental setting.  Three main experiments have been performed: (a) absence of feedback; (b) moderate feedback; and (c) intense feedback.  Several important properties of devices and circuits were inferred from measurements of current gain, voltage gain, input and output resistance, Early voltage and THD.  These parameters are directly related to the performance of the circuit regarding several of its properties, including linearity.  The first surprising experimental result was obtained by considering the distribution of such measurements individually, in terms of averages and variations, as well as jointly, by using PCA.  It has been shown, for the no-feedback case, that the great majority of the several (15) transistors types have well-defined features that discriminate them into clearly segregated groups.  This tends to justify the existence of so many different transistors types, even in a same application niche.  The incorporation of moderate feedback tended to bring the transistors groups closer one another, but good separation was preserved among several of the considered types.  However, the effect on linearization was moderate, corresponding to an improvement of about half in the respective THDs.  The implementation of intense negative feedback globally reduced the separation between transistors groups, confirming the ability of such a mechanism to promote device parameter invariance.   However, several groups remained separated, suggesting that even intense negative feedback is unable to provide perfect device invariance.   This suggests that, even in presence of intense negative feedback, the circuit will still preserve a ``memory" of certain device properties.  Therefore, it could be interesting to consider the properties of types of transistors (or even of individual devices) at the design stage.   In addition, the normalization induced by feedback took place with varying efficiency regarding the several considered measurements.  Interestingly, the intensification of negative feedback did not tend to enhance the linearity much further.  Complementary findings were provided by the subsequent analysis of pairwise measurements and, especially, by deriving a graph representation from such relationships.  In addition to confirming many of the above conclusions, such an investigation allowed the more in depth understanding of how the measurements are related in the three considered experiments. 

In addition to the results and findings summarized above, the current work also provides an objective revision of the several involved aspects and includes the analytical derivation, through a small signal hybrid model, of the main circuit properties and sensitivities, which helped in the investigation and interpretation of the obtained results.

It should be reminded that the results presented in this work are specific to the considered devices, circuit types and configurations.  The extension of such findings to other situations is not immediate and constitutes subject of further related investigations. At the same time, the significance of the obtained results can be readily gauged by the many derived prospects for future investigation.  For instance, the identification of consistent, discriminating properties of transistors types and families justifies the application of the adopted, as well as additional, multivariate statistics and pattern recognition methods for the analysis of more types of devices and circuit configurations.  Not less important is the fact that even very intense feedback levels can be unable to erase the ``memory" of the circuit with respect to some of the device parameters.   Therefore, it could be interesting to consider specific parameters of transistors and other devices during the design and implementation stages of several linear circuits.  The great complexity of interdependence between circuit parameters and components, not to mention the varying effects of negative feedback depending on the type of non-linearity, also motivates the development and adoption of complex systems and artificial intelligence concepts and methods for design and analysis of linear systems.  Another intriguing possibility which is being developed by the authors is to investigate situations analogue to those reported here  with respect to other areas, such as complex networks~\cite{newman2010networks,rodrigues2009signal}.  More specifically, it would be interesting to investigate how negative (and positive) feedback could affect the interdependence between complex networks parameters (including topology) and dynamical behaviour.

\section*{Acknowledgments}

L. da F. Costa thanks CNPq (Grant no. 307333/2013-2) for support. F. N. Silva acknowledges FAPESP (Grant No. 15/08003-4). C. H. Comin thanks FAPESP (Grant No. 15/18942-8) for financial support. This work has been supported also by FAPESP grant 11/50761-2.

\section*{Appendix A - Theoretical analysis of negative feedback }

\begin{figure*}[h]
  \begin{center}
  \includegraphics[width=0.9\linewidth]{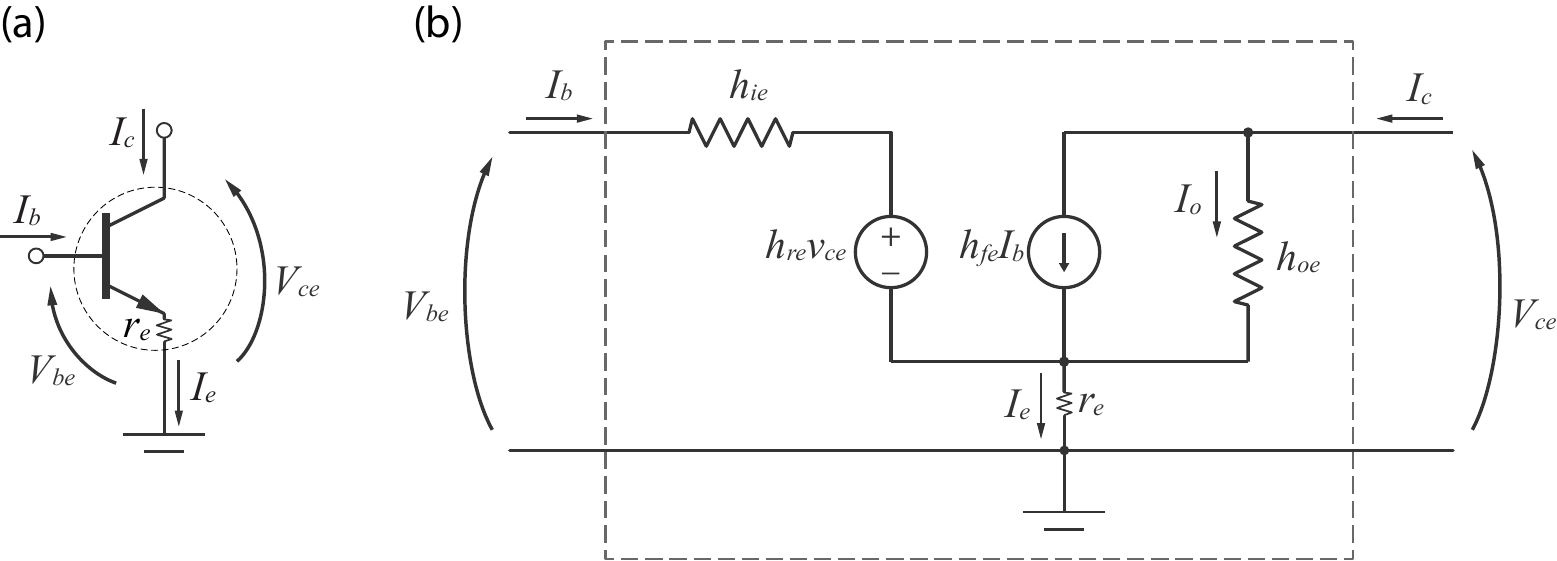} \\
  \caption{(a) Typical parameters used for describing the operation of a transistor. (b) hybrid h-parameter model of a transistor.}
  ~\label{f:portModelTransistor}
  \end{center}
\end{figure*}

\begin{figure*}[h]
  \begin{center}
  \includegraphics[width=0.7\linewidth]{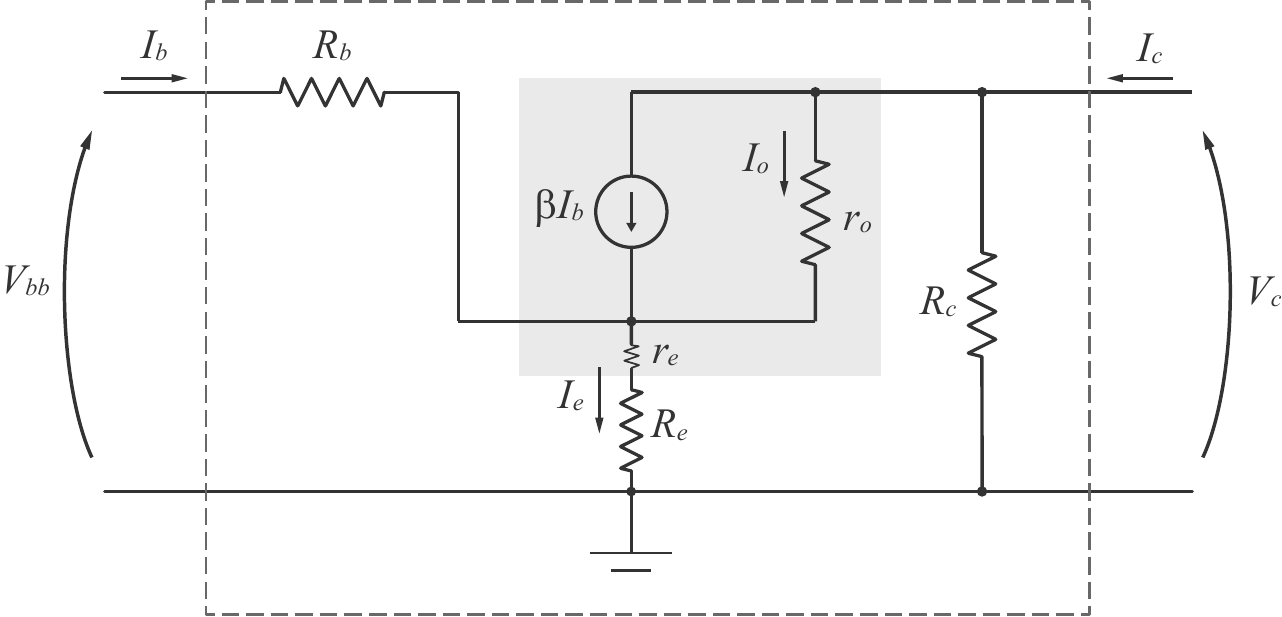} \\
  \caption{Hybrid h-parameter model of the circuit considered in the current work (shown in Figure~\ref{f:amplifierWithFeedback}). The shaded region represents the h-parameter model of the transistor.}
  ~\label{f:portModelCircuit}
  \end{center}
\end{figure*}

The experimental circuit shown in Figure~\ref{f:amplifierWithFeedback} can be analyzed 
using a small signal hybrid h-parameter model~\cite{chen2016active}. This is done by first considering that the properties of a single transistor, shown in Figure~\ref{f:portModelTransistor}(a), 
can be cast into an equivalent two-port network containing appropriate h-parameters, as shown in Figure~\ref{f:portModelTransistor}(b). 
Likewise, we can represent the circuit used in the experiments (shown in Figure~\ref{f:amplifierWithFeedback}) by a similar h-parameter circuit, which is shown in Figure~\ref{f:portModelCircuit}. Note that we considered $h_{ie}=0$ in the equivalent circuit. Using Kirchhoff's circuit laws, the following system of equations can be defined for the equivalent circuit

\begin{align}[left=\empheqlbrace]
I_c & = I_o + \beta I_b \label{eq:AppIc}\\
I_e & = I_b + I_c \label{eq:AppIe}\\
V_c & = r_o I_o + R_e I_e \label{eq:AppVc}\\ 
V_{bb} & = R_b I_b + R_e I_e \label{eq:AppVbb}\\
V_c & = R_c I_c \label{eq:AppVc}
\end{align}

The voltage amplification of the circuit is given by

\begin{equation}
A_c^f=\frac{V_c}{V_{bb}}.\label{eq:AcfDefAppendix}
\end{equation}
Solving the system of equation for $V_c$ and $V_{bb}$ and replacing the result into Equation~\ref{eq:AcfDefAppendix} results in 

\begin{equation}
A_c^f =\frac{R_c(\beta r_o - R_e - r_e)}{R_b(r_o - R_c)+((1+\beta)r_o + R_b - R_c)(R_e+r_e)}. \label{eq:AcfAppendix}
\end{equation}
In the absence of feedback, $R_e=0$ and we obtain the open-loop amplification, given by

\begin{equation}
A_c =\frac{R_c(\beta r_o - r_e)}{R_b(r_o - R_c)+((1+\beta)r_o + R_b - R_c)r_e}. \label{eq:AcAppendix}
\end{equation}

Equation~\ref{eq:AcfAppendix} can be simplified if we specifically consider the parameters used in the present work. First, we rewrite the equation as

\begin{equation}
A_c^f =\frac{R_c\beta r_o(1 - \frac{R_e}{\beta r_o} - \frac{r_e}{\beta r_o})}{R_b(r_o - R_c)+\beta r_o(1+\frac{1}{\beta r_o} + \frac{R_b}{\beta r_o} - \frac{R_c}{\beta r_o})(R_e+r_e)}. \label{eq:ampCircuit}
\end{equation}
According to the values presented in Table~\ref{t:parameterValuesExperiments}, we have that, on average, $\beta r_o = 449*14822 = 6655078\,\Omega$. On the other hand, the maximum value among all considered circuit resistances and feedback situations is $M_p = \max\{R_b, R_c, R_e, r_e\} = 32640\,\Omega$. Since $\beta r_o >> M_p$, we can consider that $M_p/(\beta r_o)\approx 0$. Therefore, the amplification can be approximated as

\begin{equation}
A_c^f \approx\frac{R_c\beta r_o}{R_b(r_o - R_c)+\beta r_o(R_e+r_e)}. \label{eq:AcfApproxAppendix}
\end{equation}

It is also possible to associate Equation~\ref{eq:AcfAppendix} with the general expression of a feedback circuit, provided by Equation~\ref{eq:ampGeneralFeedback}. This is done by replacing $A^f$ and $A$ in Equation~\ref{eq:ampGeneralFeedback} by their respective values, $A_c^f$ and $A_c$, for the considered circuit:

\begin{equation}
A_c^f = \frac{A_c}{1+A_cf}.\label{eq:ampasdf}
\end{equation}
Thus

\begin{equation}
f = \frac{A_c-A_c^f}{A_cA_c^f}.\label{eq:ampqwer}
\end{equation}
Replacing Equations~\ref{eq:AcfAppendix} and~\ref{eq:AcAppendix} into Equation~\ref{eq:ampqwer} gives

\begin{equation}
f = \frac{R_e \left(R_b+\beta  r_o\right) \left((\beta +1) r_o-R_c\right)}{R_c \left(r_e-\beta  r_o\right)\left(r_e+R_e-\beta  r_o\right)}.\label{eq:fAppendix}
\end{equation}
This equation can be approximated through a similar procedure used for deriving Equation~\ref{eq:AcfApproxAppendix}. First, Equation~\ref{eq:fAppendix} is rewritten as

\begin{equation}
f = \frac{R_e \left(\frac{R_b}{\beta r_o}+1\right) \left(1+\frac{1}{\beta r_o}-\frac{R_c}{\beta r_o}\right)}{R_c \left(\frac{r_e}{\beta r_o}-1\right)\left(\frac{r_e}{\beta r_o}+\frac{R_e}{\beta r_o}-1\right)}.
\end{equation}
Since $\beta r_o$ is significantly larger than all other parameters, we obtain

\begin{equation}
f \approx \frac{R_e}{R_c}.
\end{equation}

Another interesting quantity that can be calculated is the variation of amplification $A_c^f$ respective to a variation of the current gain $\beta$ of the transistor. This is done by calculating the derivative of Equation~\ref{eq:AcfAppendix} with respect to $\beta$, resulting in
\begin{equation}
\frac{\partial A_c^f}{\partial\beta} = \frac{r_oR_c(R_b+R_e+r_e)(r_o-R_c+R_e+r_e)}{{\mathcal{D}}^2},
\end{equation}
where $\mathcal{D} = R_b(r_o - R_c)+((1+\beta)r_o + R_b - R_c)(R_e+r_e)$. The sensitivity $S_{A_c^f}(\beta)$  of $A_c^f$ with respect to $\beta$ is defined as~\cite{chen2016active}

\begin{equation}
S_{A_c^f}(\beta) \equiv \frac{\beta}{A_c^f}\frac{\partial A_c^f}{\partial\beta},
\end{equation}
Thus

\begin{equation}
S_{A_c^f}(\beta) = \frac{\beta  r_o \left(R_b+R_e+r_e\right) \left(r_o-R_c+R_e+r_e\right)}
       { \mathcal{D}\left(\beta r_o -R_e-r_e\right)}.\label{eq:SAcfVsBetaAppendix}
\end{equation}
Since $\beta r_o$ is large, the sensitivity can be approximated as

\begin{equation}
S_{A_c^f}(\beta)\approx 
\frac{\left(R_b+R_e+r_e\right) \left(r_o-R_c+R_e+r_e\right)}
     {R_b \left(r_o-R_c\right)+\beta r_o(R_e + r_e)}.
\end{equation}

The variation of the amplification with respect to $r_o$ can also be calculated:
\begin{align}
& \frac{\partial A_c^f}{\partial r_o} = \nonumber\\
& = \frac{R_c \left(R_b+r_e+R_e\right) \left(-\beta  R_c+(\beta +1)(R_e + r_e)\right)}
     {\mathcal{D}^2}
\end{align}
The respective sensitivity of $A_c^f$ with respect to variations of $r_o$ is given by

\begin{equation}
S_{A_c^f}(r_o) \equiv \frac{r_o}{A_c^f}\frac{\partial A_c^f}{\partial r_o}
\end{equation}
Thus

\begin{align}
& S_{A_c^f}(r_o) = \nonumber\\
& = \frac{r_o \left(R_b+r_e+R_e\right) \left(-\beta  R_c+(\beta +1) (R_e+r_e)\right)}
      { \mathcal{D}\left(\beta r_o - r_e - R_e\right)},\label{eq:SAcfVsRoAppendix}
\end{align}
which can be approximated as

\begin{equation}
S_{A_c^f}(r_o) \approx \frac{\left(R_b+r_e+R_e\right) \left(-R_c+r_e+R_e\right)}
      {R_b \left(r_o-R_c\right)+\beta r_o(R_e+r_e)}
\end{equation}

We can also derive an expression for the sensitivity of amplification $A_c^f$ to small variations in open loop amplification $A_c$. This is done by calculating the derivative of $A_c^f$ with respect to $A_c$ using Equation~\ref{eq:ampasdf}:

\begin{align}
& \frac{\partial A_c^f}{\partial A_c} = \frac{1}{(1+fA_c)^2} = \nonumber\\
& \frac{1}{\left(\frac{R_e \left(R_b+\beta  r_o\right) \left((\beta +1) r_o-R_c\right)}{\left(r_e+R_e-\beta r_o\right) \left(r_e \left(R_b-R_c+(\beta +1) r_o\right)+R_b \left(r_o-R_c\right)\right)}-1\right){}^2}
\end{align}
Note that this derivative can also be written as

\begin{equation}
\frac{\partial A_c^f}{\partial A_c} = \left(\frac{A_c^f}{A_c}\right)^2.
\end{equation}
Therefore, the sensitivity of $A_c^f$ with respect to $A_c$ is given by

\begin{align}
& S_{A_c^f}(A_c) = \frac{A_c}{A_c^f}\frac{\partial A_c^f}{\partial A_c} = \frac{A_c^f}{A_c} = \nonumber\\
& \frac{\left(r_e+R_e-\beta  r_o\right) \left(r_e \left(R_b-R_c+(\beta +1) r_o\right)+R_b \left(r_o-R_c\right)\right)}
  {\left(r_e-\beta  r_o\right)\mathcal{D}}
\end{align}
Considering again the fact that $\beta r_o$ is much larger than other terms, we have that

\begin{equation}
S_{A_c^f}(A_c) \approx
\frac{r_e \beta r_o+R_b \left(r_o-R_c\right)}
     {\beta r_o(R_e + r_e)+R_b \left(r_o-R_c\right)}.
\end{equation}

The output resistance of the circuit, $R_{o}=V_c/I_c$ can be found by setting $V_{bb}=0$ in the equivalent h-parameter model of the circuit. In other words, the input becomes short-circuited. Equations~\ref{eq:AppIc} to~\ref{eq:AppVbb} can then be used to find
\begin{equation}
R_{o} = \frac{R_b r_o + (R_b + (1+\beta)r_o)(R_e+r_e)}{R_b+R_e+r_e}.
\end{equation}
This equation can be rewritten as

\begin{equation}
R_{o} = \frac{R_b r_o + \beta r_o(\frac{R_b}{\beta r_o} + \frac{1}{\beta r_o}+1)(R_e+r_e)}{R_b+R_e+r_e}.
\end{equation}
If $\beta r_o>>R_b$, we have that

\begin{equation}
R_{o} \approx r_o\frac{R_b + \beta R_e + \beta r_e}{R_b+R_e+r_e}
\end{equation}

In a similar fashion, the input resistance, $R_{i}=V_{bb}/I_b$, is found by setting $V_c=0$ and solving the system of equations for $V_{bb}$ and $I_b$, giving

\begin{equation}
R_{i} = \frac{R_b r_o + (R_b + (1+\beta)r_o)(R_e+r_e)}{R_e+r_e+r_o}.
\end{equation}

\begin{equation}
R_{i} \approx r_o \frac{R_b + \beta R_e+\beta r_e}{R_e+r_e+r_o}.
\end{equation}

\section*{Appendix B - Transistor parameters estimation method}

Finding a representative and practical estimation of the transistor parameters ($R_o$, $\beta$ and $V_a$) is an important task. While some datasheets traditionally present the transistors parameters in terms of average, maximum and minimum, the transistor parameters can vary substantially depending on the circuit configuration space ($I_c \times V_c \times V_{bb}$)~\cite{silva2016seeking}, which complicates the estimation of the parameters. We developed a method to estimate the representative parameters of a transistor, in which we employed the Early Voltage (see Section~\ref{s:EarlyExplanation}) as a measurement of configuration stability, i.e. we try to identify the $V_{bb}$ interval where $V_a$ is more stable. This section describes this methodology.

To estimate the early voltage, we start by considering all the isolines (corresponding to constant $V_{bb}$) obtained experimentally for a transistor. For each isoline, the Early voltage $V_a(V_{bb})$ is estimated by applying Least Squares Method~\cite{draper1966applied} along its last $100$ data points. This constraint is necessary in order to avoid the non-linear saturation region. Because this estimation is still susceptive to noise, we improve the signal-noise ratio of $V_a(V_{bb})$ by applying the Savitzky–Golay filter (S-G)~\cite{savitzky1964smoothing}, which approximates the data to a set of smooth polynomial functions.

\begin{figure*}[h]
  \begin{center}
  \includegraphics[width=0.95\linewidth]{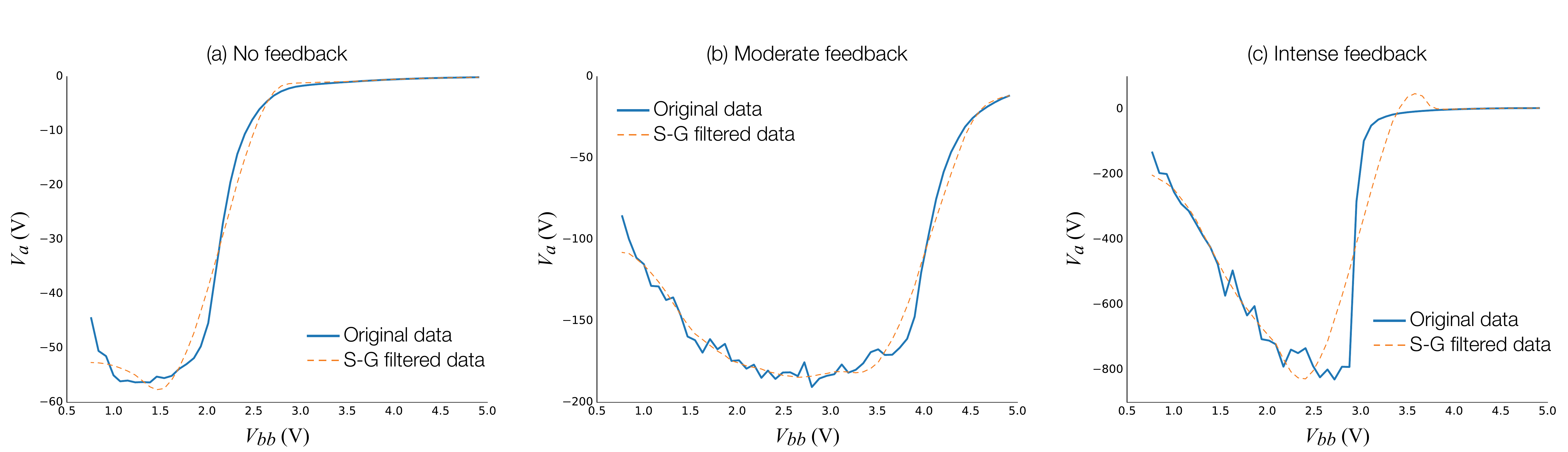} \\
  \caption{Three examples, respective to each of the performed experiments --- (a) no feedback, (b) moderate feedback, and (c) intense feedback, of experimental curves of $V_a$ in terms of $V_{bb}$.}
  ~\label{f:VEarlyExamples}
  \end{center}
\end{figure*}

Figure~\ref{f:VEarlyExamples} shows the curves of $V_a(V_{bb})$ estimated for a transistor under the three feedback conditions considered in this work: no feedback (a), moderate (b) and intense feedback (c). In all cases, the curves start with a prominent depression region followed by a fast increase of $V_a$ with $V_{bb}$. Interestingly, this result holds for all the considered transistors, but the minimum attained value $V_{a, \min}$ and the corresponding isoline $V_{bb, \min}$ can vary from case to case. Since the depression region indicates a region of well-behaved $V_a$, we use the minimum of the curve, $V_{a, \min}$, as a representative value for the early voltage.

\begin{figure*}[h]
  \begin{center}
  \includegraphics[width=0.95\linewidth]{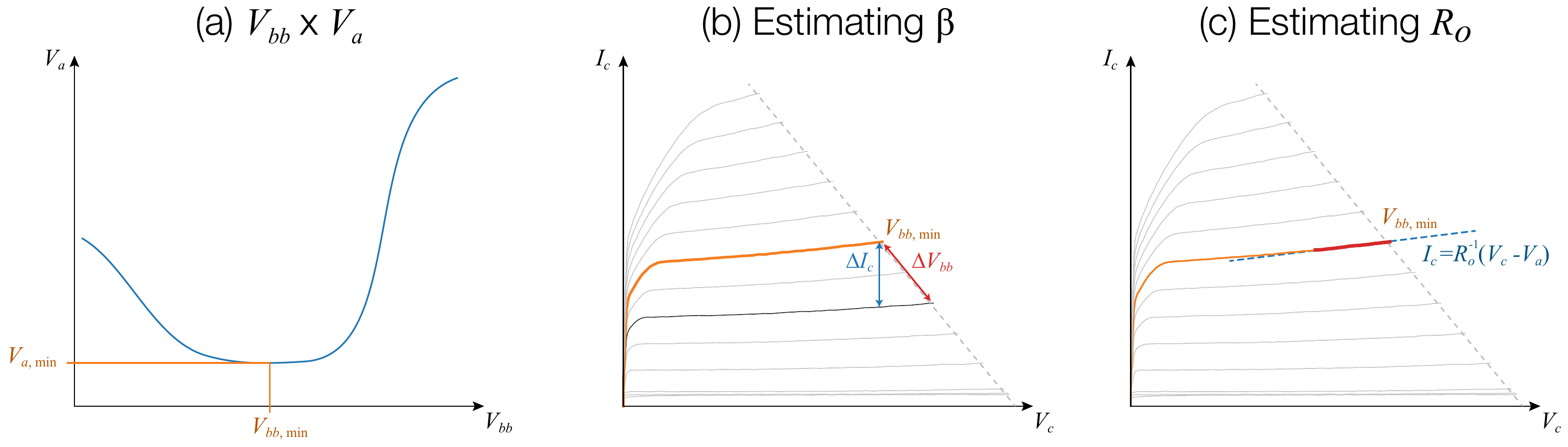} \\
  \caption{(a) A typical curve $V_a$ in terms of $V_{bb}$, with the respectively obtained values $V_{a,\min}$ and $V_{bb,\min}$. (b) The space $I_c$ in terms of $V_c$ and some of the isolines parametrized by $V_{bb}$, including that corresponding to $V{bb,\min}$, as well as the respective $\Delta V_{bb}$ and $\Delta I_{c}$. (c) The same isolines as in (b), but with the setting necessary to estimate $R_o$. }
  ~\label{f:parameterEstimation}
  \end{center}
\end{figure*}

Having defined a stable reference for $V_a$ (as shown in Figure~\ref{f:parameterEstimation}(a)), a working interval can be imposed upon $V_{bb}$ and $I_c$, spanning from $V_{bb, \min}$ to two previous isolines (the experiments always take 64 isolines corresponding to a $V_{bb}$ step of $5/64V$), as indicated in Figure~\ref{f:parameterEstimation}(b).  Now, $\beta$ can be estimated as $\Delta I_c / \Delta I_b$, where $\Delta I_b = \Delta V_{bb} / R_b$.  This approach assumes small $V_b$ variation along the working interval.  The estimation of $R_o$ can be performed by numerically calculating (by using least square approximation) the slope of the $V_{bb,\min}$ isoline considering its last 100 points, as illustrated in Figure~\ref{f:parameterEstimation}(c).

\bibliographystyle{unsrt}
\bibliography{references}

\end{document}